\documentclass{aa}

\usepackage{times}
\usepackage{graphicx}
\usepackage{xspace}
\usepackage{epsfig}
\usepackage{natbib}
\usepackage{rotating}
\usepackage{dcolumn}

\newcommand{\loglhalbol}{\ensuremath{\log{(L_{\rm H\alpha}/L_{\rm bol})}}\xspace}
\newcommand{\denis}{DENIS\,1048-3956\xspace}
\newcommand{\denisstar}{DENIS\,1228-1547\xspace}
\newcommand{\pcstar}{PC\,0025+0447\xspace}
\newcommand{\tvlm}{TVLM\,513-46546\xspace}
\newcommand{\bri}{BRI\,B0021-0214\xspace}
\newcommand{\lsr}{LSR\,J1835+3259\xspace}
\newcommand{\tm}{2MASS\,J0036159+182110\xspace}

\begin{document}

\title{The ultracool dwarf DENIS-P\,J104814.7-395606}

\subtitle{Chromospheres and coronae at the low-mass end of the main-sequence}

\author{B. Stelzer \inst{1} \and J. Alcal\'a \inst{2} \and K. Biazzo \inst{2} \and 
B. Ercolano \inst{3} I. Crespo-Chac\'on \inst{4} \and J. L\'opez-Santiago \inst{4} \and 
R. Mart\'inez-Arn\'aiz \inst{4} \and J. H. M. M. Schmitt \inst{5} \and 
E. Rigliaco \inst{6} \and F. Leone \inst{7} \and G. Cupani \inst{8}}

\offprints{B. Stelzer}

\institute{INAF - Osservatorio Astronomico di Palermo,
  Piazza del Parlamento 1,
  90134 Palermo, Italy \\ \email{B. Stelzer, stelzer@astropa.inaf.it} \and
  INAF - Osservatorio Astronomico Capodimonte, Salita Moiariello 16, 80131 Napoli, Italy \and
  Universit\"atssternwarte M\"unchen, Scheinerstrasse 1, 81679 M\"unchen, Germany \and
Dpto. de Astrof\'isica y Ciencias de la Atm\'osfera, Universidad Complutense de Madrid, 28040 Madrid, Spain \and
  Hamburger Sternwarte, Gojenbergsweg 112, 21029 Hamburg, Germany \and 
  Lunar and Planetary Laboratory, University of Arizona, 1629 E. University Blvd, Tucson, AZ 85721, USA \and
  INAF - Osservatorio Astronomico di Catania, Via S.Sofia 78, 95123 Catania, Italy \and 
  INAF - Osservatorio Astronomico di Trieste, Via G.B. Tiepolo 11, 34143 Trieste, Italy 
} 

\titlerunning{Chromosphere and corona of DENIS-P\,J104814.7-395606}

\date{Received $<$15 September 2011$>$ / Accepted $<$22 November 2011$>$}

\abstract{Several diagnostics ranging from the radio to the X-ray band are suitable for 
investigating the magnetic activity of late-type stars. 
Empirical connections between the emission at different wavelengths 
place constraints on the nature and efficiency of the emission mechanism 
and the physical conditions in different atmospheric layers. 
The activity of ultracool dwarfs, at the low-mass end of the main-sequence,  
is poorly understood. 
}
{We have performed a multi-wavelength study of one of the nearest
M9 dwarfs, DENIS-P\,J104814.7-395606 ($4$\,pc), 
with the aim to examine its position
within the group of magnetically active ultracool dwarfs, and, in general, to advance 
our understanding of these objects comparing them to early-M type dwarf stars and the Sun. 
}
{We have obtained an XMM-Newton observation of DENIS-P\,J104814.7-395606 and a broad-band
spectrum from the ultraviolet to the near infrared with X-Shooter. From this dataset 
we obtain the X-ray properties, stellar parameters, kinematics and the emission line
spectrum tracing chromospheric activity. 
We integrate these data by a compilation of activity parameters for 
ultracool dwarfs from the literature. 
}
{Our deep XMM-Newton observation has led to the first X-ray detection of 
DENIS-P\,J104814.7-395606
($\log{L_{\rm x}} = 25.1$) 
as well as the first measurement of its $V$ band brightness ($V = 17.35$\,mag). 
Flux-flux relations 
between X-ray and chromospheric activity indicators are here for the first time
extended into the regime of the ultracool dwarfs.
The approximate agreement of DENIS-P\,J104814.7-395606 
and other ultracool dwarfs 
with flux-flux relations for early-M dwarfs suggests that the 
same heating mechanisms work in the atmospheres of ultracool dwarfs, 
albeit weaker as judged from their lower fluxes. 
The observed Balmer decrements of \denis are compatible with optically thick plasma
in LTE at low, nearly photospheric temperature or optically thin LTE plasma at
$20000$\,K. 
Describing the decrements with Case\,B recombination requires different emitting
regions for H$\alpha$ and the higher Balmer lines.
The high observed H$\alpha$/H$\beta$ flux ratio is also poorly fitted by the 
optically thin models. 
We derive a similarly high value for the H$\alpha$/H$\beta$ ratio of vB\,10 and 
LHS\,2065 and conclude that this may be a characteristic of ultracool dwarfs. 
We add DENIS-P\,J104814.7-395606 to the list of ultracool dwarfs detected in both the 
radio and the X-ray band. 
The Benz-G\"udel relation between radio and X-ray luminosity of late-type stars is 
well-known to be violated by ultracool dwarfs. 
We speculate on the presence of two types of ultracool dwarfs
with distinct radio and X-ray behavior.}
{}


\maketitle

\section{Introduction}\label{sect:intro}

Late-type stars have long been known to display signatures of magnetic activity
evidencing solar-like dynamo action \citep{Rosner85.1}. 
Diagnostics of activity are available across all the electromagnetic spectrum of such stars. 
The various indicators of magnetic activity 
probe different layers of the atmosphere. 
In the radio band, gyrosynchrotron emission from electrons moving in the coronal 
magnetic field is observed \citep{Guedel94.1}. The optical regime is characterized by 
chromospheric emission lines \citep{Hawley96.1}, 
and the X-ray band displays thermal emission from heated plasma in coronal loops
\citep{Schmitt95.1}. 
Strong variability is typical for all radiation originating from magnetic activity, and 
flares 
belong to its most obvious manifestations. 

In particular, in dM(e) stars (``flare stars'') the activity is ubiquitous
and flares are more frequently observed 
than on the Sun.  
What happens at and beyond 
the cool end of the main-sequence (spectral types from late-M to L)  is less clear. 
Very low-mass (VLM) stars and brown dwarfs (BDs) are thought to be
fully convective and hence a solar-type dynamo is not expected to work. 
Throughout this paper 
we refer to objects with spectral type equal to or later than M7 as ultracool dwarfs (UCDs), 
regardless of whether they are VLM stars or BDs.

Various optical/infrared emission lines, such as the Balmer series, \ion{Ca}{ii}\,H\&K,
\ion{Ca}\,Infrared Triplet (\ion{Ca}{ii}\,IRT), and several He\,I lines, trace the 
chromospheric activity of late-type stars. Empirical relations between the fluxes emitted in
these diagnostics yield information on the physical state and origin of the emitting plasma, and their
dependence on stellar parameters such as spectral type allow to unveil how universal the
properties of chromospheres are. Such flux-flux relations have to our knowledge never been
examined for UCDs where H$\alpha$ emission is usually the only available observable of 
chromospheric activity. 

Surveys of chromospheric H$\alpha$ activity in UCDs in the solar neighborhood
\citep[e.g.][]{Gizis00.1, Reid02.2, Mohanty03.1, Schmidt07.1}
have shown that H$\alpha$ emission reaches a maximum at
spectral type M7, and fades off for the latest M dwarfs.
The X-ray regime is widely unexplored beyond spectral type M6.
Generally, only upper limits for the X-ray luminosity ($L_{\rm x}$) are available for a small number of
UCDs in the field, and are quite ill-constrained. 
A possible explanation for the decline of the H$\alpha$ activity
was put forth by \cite{Mohanty02.1}
who suggested that chromospheric emission might be suppressed
because the high electrical resistivity in the cool, neutral atmosphere prevents
build-up of substantial magnetic stress. 
Similar arguments might explain the apparently weak X-ray coronae of UCDs.
Direct magnetic field measurements by means of 
near-infrared (NIR) spectroscopy have demonstrated the presence of kilo-Gauss fields on 
UCDs \citep{Reiners09.3} but it is unclear at present if these fields have a structure
that is adequate for storing energy.
On the other hand, UCDs seem to be overluminous in radio by factors
of $> 1000$ with respect to the
$L_{\rm R}$ vs. $L_{\rm X}$ relation observed for earlier-type active stars 
\citep{Berger05.1}. 
The detection of radio emission indicates a high-density electron population and the 
presence of magnetic fields. 

A handful of M8 to early-L dwarfs in the solar neighborhood has been found to
undergo extraordinary strong H$\alpha$, radio or X-ray emission
\citep[for references see][]{Liebert03.1, Berger06.1, Stelzer06.1}. 
In most of these cases the emission seems to be due to flares.
Studies of UCD radio properties showed in some objects 
the presence of a highly circularly polarized pulsed emission component  
overlaid on their quiescent, non-variable (gyro-synchrotron) emission 
\citep[e.g.][]{Burgasser05.1, Hallinan06.1, Hallinan08.1}. 
The periodicities of the spikes are 
consistent with the rotational velocities of the objects implying an origin in a 
beaming mechanism. This radio component has been attributed to 
the electron cyclotron maser (ECM) instability, a phenomenon 
seen in solar system giant planets rather than in stars. 

Seeking to provide further constraints on the radio/H$\alpha$/X-ray connection
of UCDs, we have observed DENIS-P\,J104814.7-395606 (henceforth \denis) 
with {\em XMM-Newton}. \denis is one of the nearest 
UCD \citep[$4.00 \pm 0.03$\,pc; ][]{Costa05.0}. 
It was classified as an old M9 dwarf because of the
absence of lithium absorption \citep{Delfosse01.1}. A huge chromospheric flare,
identified in a sequence of four high-resolution spectra, 
has been reported by \cite{Fuhrmeister04.1}. Mass motions related to the flare
were inferred from the detection of blushifts 
in some emission lines, and interpreted as rising gas cloud. 
Radio bursts were observed on \denis indicative of ECM emission \citep{Burgasser05.1}, 
but its radio spectrum shows a negative slope typical 
for gyrosynchrotron emission \citep{Ravi11.0}.
No X-ray emission could be detected with {\em ROSAT} \citep{Schmitt04.1}. 

Here we present a deeper X-ray image of \denis obtained with {\em XMM-Newton}
and a new broad-band spectrum covering the full optical range and
the NIR obtained with X-Shooter at the VLT. 
The data analysis and results from these two observations are presented in 
Sects.~\ref{sect:data_xmm} and~\ref{sect:data_xshooter}, respectively. 
In Sect.~\ref{sect:multi_lambda}
we put our findings in context of magnetic activity on other UCDs and earlier-type M dwarfs. 
We present for the first time flux-flux relations between chromospheric and coronal
radiation involving UCDs.
We compare the observed Balmer decrements of \denis to those of other late-type 
dwarf stars and to theoretical predictions
from which we obtain clues on the physical conditions of the emitting region.
Finally, we discuss the connection between radio and X-ray emission of UCDs 
including the new data for \denis. 
All this is achieved by making use of an 
updated compilation of the X-ray, H$\alpha$ and radio properties of UCDs. We have also 
analysed archived and as yet unpublished X-ray data for two UCDs, 
DENIS-P\,J1228.2-1547 (henceforth abbreviated \denisstar) and \pcstar. 
A summary of our results and conclusions are given in Sect.~\ref{sect:summary}.

\begin{table*}[t]
\begin{center}
\caption{Summary of OM observations and results for \denis.}
\label{tab:om_results}
\begin{tabular}{lccccc}\hline
         & \multicolumn{5}{c}{------------------------------ Exposure number -----------------------------}  \\ 
         &  006             &  007             &  008             &  009             &  011 \\ \hline
Date-Obs [JD-2455214] & $0.025463$ & $0.067014$      & $0.107639$       & $0.184028$       & $0.221065$ \\
Expo [s]       & $4400$           & $4380$           & $4200$           & $3000$           & $2200$           \\
$\alpha_{\rm 2000}$\,[hh:mm:ss.ss]  &  10:48:13.57     &  10:48:13.58      &  10:48:13.58      & 10:48:13.56       &  10:48:13.55 \\
$\delta_{\rm 2000}$\,[dd:mm:ss.ss]  & $-$39:56:16.87   & $-$39:56:16.85    & $-$39:56:16.90    & $-$39:56:16.82    & $-$39:56:16.89 \\
$V_{\rm ins}$ [mag] & $17.32 \pm 0.03$ & $17.26 \pm 0.03$ & $17.37 \pm 0.03$ & $17.42 \pm 0.04$ & $17.40 \pm 0.04$ \\
\hline
\end{tabular}
\end{center}
\end{table*}

\section{XMM-Newton data analysis}\label{sect:data_xmm}

\denis was observed by {\em XMM-Newton} for $20$\,ksec on Jan 17, 2010
(Obs-ID\,0600410101). 
{\em XMM-Newton} acquired both X-ray and optical data for \denis using EPIC and the Optical
Monitor (OM). These 
data were analysed with Standard {\em XMM-Newton} Science Analysis System (SAS) software (v9.0).

\subsection{X-ray data}\label{subsect:data_xray}

This data analysis closely follows the procedure that was described in detail by 
\cite{Barrado11.0} and we briefly summarize here the individual steps. 
We first created photon event lists with the metatasks {\sc epchain} and {\sc emchain} 
for EPIC/pn and EPIC/MOS respectively. After standard data filtering we performed
source detection for each of the three detectors. \denis is detected as a weak
X-ray source on EPIC/pn but not on the two MOS detectors. 
Most of the photons detected in EPIC/pn have energies $<2$\,keV. 
Therefore, to maximize the sensitivity, 
in a second step we merged the data of all three instruments and then
repeated the source detection procedure on the merged data for energies 
between $0.2$ and $2.0$\,keV. 
Taking account of the
different relative sensitivity of EPIC/pn and EPIC/MOS by scaling the exposure maps 
of EPIC/pn accordingly, 
the resulting combined data set has an effective EPIC/MOS exposure of $101$\,ksec
at the position of \denis. 
We then repeated the source detection procedure on the merged data, and found a
source count rate of $(1.16 \pm 0.14)\,10^{-3}$\,cts/s for \denis. 

Due to the low photon statistics we attempt no spectral analysis and assume 
a plasma temperature between $0.6...1.0$\,keV to estimate the flux
with help of 
PIMMS\footnote{The Portable Interactive Multi-Mission Simulator is available at 
http://cxc.harvard.edu/toolkit/pimms.jsp}.  
This yields an intrinsic X-ray luminosity of $\log{L_{\rm x}}\,{\rm [erg/s]} = 25.1$
with practically no dependence on the temperature within the assumed range.
Our measurement is consistent with the upper limit measured from {\em ROSAT} data 
\cite[$\log{L_{\rm x}}\,{\rm [erg/s]} < 26.3$;][]{Schmitt04.1}. 
\cite{Fuhrmeister04.1} give a bolometric luminosity of $1.5 \cdot 10^{30}$\,erg/s
using $d = 4.6$\,pc. Correcting this value for our updated distance estimate, 
the fractional X-ray luminosity of \denis is 
$\log{(L_{\rm x}/L_{\rm bol})} = -5.0$. 

We extracted a lightcurve for \denis 
and searched for variability with a maximum likelihood (ML) method that divides
the sequence of photons in intervals of constant signal 
and subtracts the background from the photon time series on a statistical basis 
\citep[see ][]{Stelzer07.1}. 
Two short and weak flare-like intensity enhancements are seen in the binned
lightcurve. Only one of these events is recognized by the ML algorithm at $95$\,\%
significance.

\subsection{Optical data}\label{subsect:data_om}

During the X-ray observation the Optical Monitor (OM) was operating in the 
{\sc Fast Mode} with the $V$ band filter. 
The central {\sc Fast Mode} window has been used with a time-resolution of $10$\,sec. 
A larger image window recorded contemporaneously with the {\sc Fast Mode} data an 
image of the region around \denis. 
The maximum allowed exposure time for the chosen {\sc Fast Mode} setup is $4400$\,ksec. 
Therefore, 
the OM observation was divided into five 
consecutive exposures (see Table~\ref{tab:om_results}). 
The OM data was processed with standard SAS procedures. 

There is no obvious variability in the $10$\,s-binned lightcurves.
However, a comparison of the average brightness levels of the five exposures 
shows small variations with minimum count rate of $1.45 \pm 0.05$\,cts/s and maximum
count rate of $1.65 \pm 0.04$\,cts/s for \denis, i.e. variations 
at a $2\,\sigma$ level on timescales of a few hours. 

The time-averaged instrumental magnitude of \denis derived from the {\sc Fast Mode}
data is $V_{\rm ins} = 17.35 \pm 0.02$\,mag.
This number can not be converted to standard magnitude with the SAS procedures 
because observations were performed in a single filter and, consequently, no color
information is available. We have estimated the conversion factor from instrumental
to standard $V_{\rm std}$ band magnitude using our previous experience with the OM. In our survey
of the young cluster Collinder\,69 \citep{Barrado11.0} $B$ and $V$ band exposures were
obtained and the (standard) $B-V$ derived for the objects in this field covers a wide range. 
For an M8 dwarf the expected\footnote{http://www.pas.rochester.edu/$\sim$emamajek/memo\_M.html}  
$B-V$ is $2.2$\,mag, and for this value we find 
from the parabolic relation between $V_{\rm ins} - V_{\rm std}$ and $(B-V)_{\rm std}$
for Collinder\,69 a correction factor of $0.11$\,mag, i.e. $V_{\rm std} = 17.24$\,mag. Combining this with
the $B$ band measurement of \denis 
from the literature we obtain $B-V = 1.76$\,mag, much smaller than
the assumed value. While this difference may be due to intrinsic source variability of the
non-contemporaneous data, there are no good constraints on 
the blue colors of late-M dwarfs. 
If we assume the lower value for $B-V$ that we derived above, the filter 
transformation factor reduces to $0.075$\,mag, and $V_{\rm std} = 17.28$\,mag. Further
iteration does not yield significant changes. Therefore, we adopt an uncertainty of 
$\pm 0.04$\,mag and a value of $V_{\rm std} = 17.35 \pm 0.04$\,mag.

The {\sc Image Mode} data do not yield additional brightness information but they can be
used for a comparison of the present-day position of \denis with its position expected
from its proper motion given in the literature. A boresight correction needs to be applied
to take account of small OM pointing errors. The OM images have artifacts such as ghosts.  
In principle, the $Q_{FLAG}$ in the source list may be used to identify reliable sources. However, 
\cite{Kuntz08.1} describe that the $SIGNIF$ parameter
is a better indicator for the reality of a source than the $Q_{FLAG}$. 
We selected by visual inspection a number of optical sources with 2\,MASS counterparts distributed
across the OM image but avoiding areas with obvious artifacts. Then, we cross-correlated the
positions of the selected objects in the two catalogs and searched iteratively for the mean
offset in $\alpha$ and $\delta$. The results are $\Delta \alpha = -0.06^{\prime\prime}$
and $\Delta \delta = -1.71^{\prime\prime}$ for Expo.No.006. These values are in the range
cited by \cite{Talavera09.1} for the typical OM pointing accuracy.  

\begin{figure*}[t]
\begin{center}
\includegraphics[width=17cm]{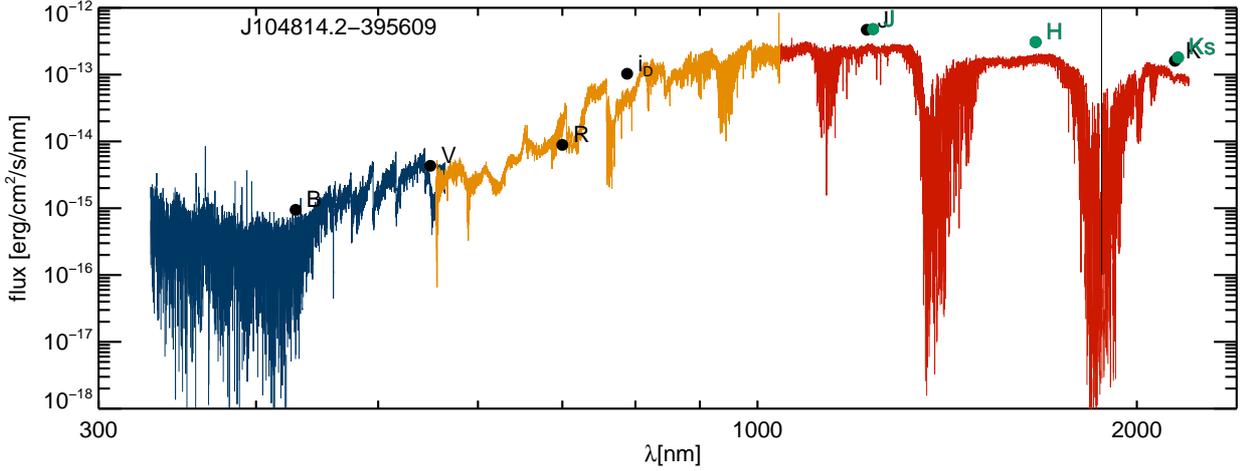}
\caption{Broad-band flux-calibrated X-Shooter spectrum of \denis. The spectra from the three
X-Shooter arms are shown in different colors. Published photometry from the literature is overplotted. 
The 2\,MASS $JHK_S$ magnitudes are shown with a slightly different color to distinguish them from the 
DENIS $JK_S$ photometry given by \cite{Delfosse01.1}. At the blue end the spectrum is dominated by noise
due to the intrinsic faintness of \denis. Telluric
absorptions are clearly visible especially in the NIR spectrum.}
\label{fig:spectrum_allarms}
\end{center}
\end{figure*}

\section{X-Shooter data analysis}\label{sect:data_xshooter}

\denis was observed on Apr 06, 2010 with the X-Shooter spectrograph at the 
VLT (ESO; Chile). The data were acquired within the INAF/GTO time \citep{Alcala11.0}.
With its three spectrograph arms, 
X-Shooter provides simultaneous wavelength coverage in the spectral range $3000-24800$\,\AA. 
Slit widths of $0.5^{\prime\prime}/0.4^{\prime\prime}/0.4^{\prime\prime}$
were used in the UVB/VIS/NIR arms, respectively,
yielding spectral resolutions of $9100/17400/11300$.
The total exposure time in each of the three spectrograph arms was $1200$\,sec. 
The data were obtained in the so-called {\em nod} mode and were reduced independently for 
 each arm with the X-Shooter pipeline v1.0.0 \citep{Modigliani10.0},
following the standard steps: bias or dark subtraction, flat fielding,
optimal extraction, wavelength calibration, and sky subtraction. 
The final outcome of this procedure is a one-dimensional spectrum that we have 
flux calibrated using a standard star observed during the same night
and corrected for atmospheric extinction with our own IDL\footnote{The Interactive Data Language (IDL) 
is a trademark of ITT Visual Information Systems.} routines.  

The flux-calibrated spectrum of all three arms is shown 
in Fig.~\ref{fig:spectrum_allarms}. At the conjunction of two arms (UVB/VIS and VIS/NIR) 
the match of the flux scale is excellent. 
Photometry from the literature
\footnote{The $BRi_DJK_S$ magnitudes are from \protect\cite{Delfosse01.1}, 
$JHK_s$ magnitude from 2\,MASS and the $V$ magnitude from our measurement with the {\em XMM-Newton} OM presented in Sect.~\ref{sect:data_xmm}. The $B$ and $R$ photometry has been converted from USNO to the standard system using the equations given on http://www.britastro.org/asteroids/USNO\%20photometry.htm.} is overlaid, providing further evidence for 
the high quality of the flux calibration. 
%
%
Note especially the very good agreement of our $V$ band
measurement from the {\em XMM-Newton} OM with the X-Shooter spectrum. 
A discrepancy of roughly a factor of two between the photometry and the X-Shooter spectrum 
is present in the NIR, possibly due to residual problems with the flux calibration of the
NIR spectrum. Note, however the good match of the NIR and the VIS spectrum. We do not know the
origin of the offset between the NIR spectrum and photometry but we do not give much weight
to it because, in practice, our data analysis concentrates on the UVB and VIS spectra.  

The spectrum of the VIS arm (550 -- 1050\,nm), properly corrected for the 
telluric absorption, was used to estimate several astrophysical 
parameters, such as spectral type, effective temperature ($T_{\rm eff}$), gravity ($\log{g}$), 
and radial velocity (RV); see Sects.~\ref{subsect:xshooter_stellarparams} 
and~\ref{subsect:xshooter_rotkine}. 
The spectra of all three arms were searched for emission lines for which we 
measured the equivalent width ($EW$) and flux ($f_{\rm line}$);
see Sect.~\ref{subsect:xshooter_emissionlines}.

\subsection{Fundamental stellar parameters}\label{subsect:xshooter_stellarparams}

\begin{figure}
\begin{center}
\vspace*{0.5cm}
\includegraphics[width=6cm, angle=270]{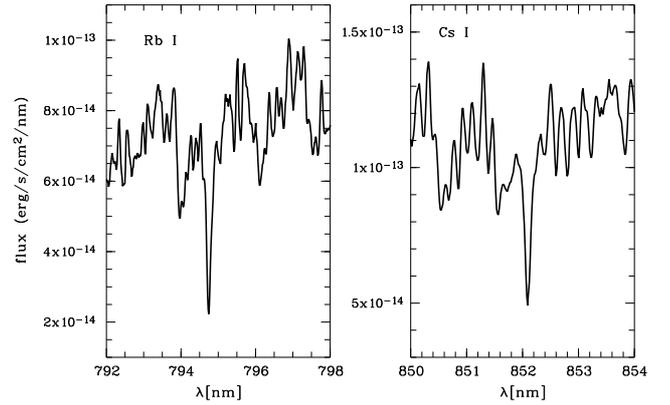}
\caption{Detail of the spectrum of \denis in the wavelength range around 
       the \ion{Rb}{i} and \ion{Cs}{i} absorption 
       lines.}
\label{fig:RbCs}
\end{center}
\end{figure}

As shown by \citet[][]{Basri00.2}, 
the \ion{Rb}{i} at $\lambda$ 794.76\,nm and \ion{Cs}{i} 
at $\lambda$ 852.11\,nm absorption lines are very sensitive to the effective 
temperature of VLM stars and BDs. These two lines are very 
well detected in our X-Shooter spectrum (see Fig.~\ref{fig:RbCs}). We measure equivalent 
widths of 0.09$\pm$0.01\,nm and 0.06$\pm$0.01\,nm for \ion{Rb}{i} and \ion{Cs}{i}, 
respectively. These values correspond to a temperature of $\approx$2450\,K 
according to the temperature scale by \citet[][]{Basri00.2}, consistent with a 
spectral type M9 and in agreement with previous values for 
the spectral type \citep[e.g.][]{Reiners10.0}. 
We have obtained an independent estimate of the spectral type using the spectral index 
PC\,3 as defined by \cite{Martin99.2} and find a spectral type of M8.8 
that matches our above results.

In order to derive other physical parameters, we use synthetic model spectra 
gathered from the {\em star, brown dwarf \& planet atmosphere web simulator}
\citep[][]{Allard10.0}, electronically 
available\footnote{http://phoenix.ens-lyon.fr/simulator/index.faces}.
A grid of synthetic spectra was generated in the parameter space 
$2300\,{\rm K} < T_{\rm eff} < 2600$\,K and $4.0<\log g<4.5$, in steps of 100\,K and 
0.5\,dex in $T_{\rm eff}$ and $\log{g}$, respectively, while keeping 
$v\sin{i}=0$ km\,s$^{-1}$ and solar metallicity fixed.
The literature values for the rotational velocity of \denis are around $20$\,km/s
but at the resolution of the X-Shooter spectrum we are not sensitive to this parameter;
see Sect.~\ref{subsect:xshooter_rotkine}. 

\begin{figure}
\begin{center}
\includegraphics[width=6.5cm, angle=270]{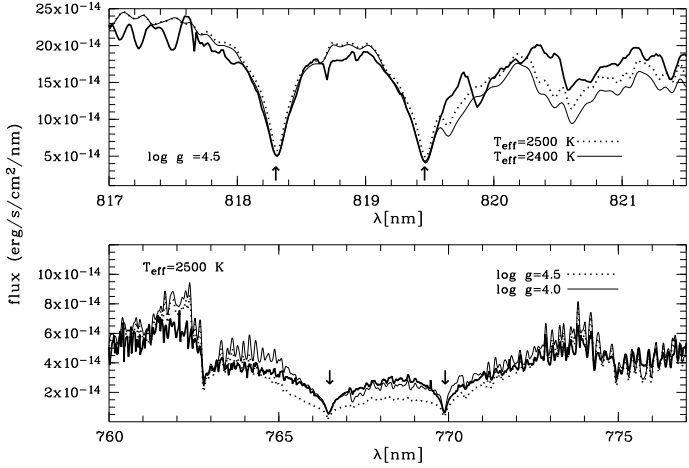}
\caption{Detail of the spectrum of \denis (thick solid lines) 
       in the wavelength range around the \ion{Na}{i} (upper panel) and
       \ion{K}{i} (lower panel) absorption doublets. The spectrum of
       \denis has been corrected for telluric absorption lines.
       The dotted line and the thin solid line represent synthetic spectra with $T_{\rm eff}$ 
       and $\log{g}$ as labelled.}
\label{fig:NaK}
\end{center}
\end{figure}

In low-mass stars and BDs the \ion{Na}{i} doublet at 
$\lambda\lambda$ 818.33, 819.48\,nm is very sensitive both to $T_{\rm eff}$ 
and $\log{g}$. These lines were used to estimate temperature and gravity. 
We find a best match between observed and synthetic spectra for 
$T_{\rm eff}=2450\pm100$\,K and $\log{g}=4.5\pm0.5$\,dex 
(see Fig.~\ref{fig:NaK}). 
The temperature is in good agreement with the value derived from the 
equivalent width of the \ion{Rb}{i} and \ion{Cs}{i} lines above. 
Fitting the \ion{K}{i} absorption doublet at $\lambda\lambda$ 766.48, 769.89\,nm 
leads to a similar temperature, but the only way to obtain a reasonable fit to 
the wings of the lines is by using a synthetic spectrum with a lower gravity.
This is shown in the bottom panel of Fig.~\ref{fig:NaK}, where the two synthetic models with 
$\log{g}=4.0$ and $\log{g}=4.5$, but with the same $T_{\rm eff}=2500$\,K,
are overplotted on the X-Shooter spectrum. A similar result is obtained when 
using the \ion{Na}{i} doublet at $\lambda\lambda$ 588.99, 589.59\,nm, although 
the S/N of the X-Shooter spectrum in this wavelength range is quite low 
($\sim$7). From the above results, we conclude that reliable values for the 
temperature and gravity of \denis are $T_{\rm eff}=2450\pm50$\,K
and $\log{g}=4.3\pm0.3$, respectively. 

\subsection{Rotation and kinematics}\label{subsect:xshooter_rotkine}

Literature values for the $v\sin{i}$ of \denis are reported to 
range from $18$ to $30$\,km/s \citep{Delfosse01.1, Fuhrmeister04.1, Reiners09.5}. 
These determinations are 
based on spectra with a much higher resolution than our X-Shooter spectrum.
Including also $v\sin{i}$ as a parameter in our fits does not 
change the results on the stellar parameters. 
We have cross-correlated the VIS X-Shooter spectrum, corrected for telluric 
contribution, with a synthetic spectrum with zero rotation. The synthetic spectrum
was retrieved from the web simulator of \cite{Allard10.0} matching both  
the resolution and astrophysical parameters of \denis. The cross-correlation 
function shows an apparently broad but well detected peak. The width 
diminuishes when strong molecular bands are not considered in the cross-correlation. 
We measured a width of $20 - 30$\,km/s depending on the placement of the cross-correlation 
background. Thus, a reasonable value for $v\,\sin{i}$ from the X-Shooter spectrum
is $25 \pm5$\,km/s, in agreement with previous measurements. 

We estimated the RV of DENIS\,1048-39 by measuring the Doppler shift of several 
\ion{Na}{i} absorption lines. The result, after application of the corresponding 
barycentric correction (-4.6\,km~s$^{-1}$) is RV$=-11.4\pm2.0$\,km~s$^{-1}$. This
result is also in agreement with previous RV determinations \citep[][]{Montes01.2}.
Using our radial velocity estimate and the proper motion components of the star, 
$\mu_{\alpha}\cos\delta$$=-1178.4\pm$12.7\,mas~yr$^{-1}$, 
$\mu_\delta$$=-986.4\pm$12.7\,mas~yr$^{-1}$, retrieved from the PPMXL Catalog 
\citep{Roeser10.1}, we can derive spatial velocity components of the star adopting 
the distance $d=4$\,pc. 
The results, corrected by solar motion \citep{Schoenrich10.0} 
in a left-handed coordinate system, are 
($U, V, W$)$=$($+0.26\pm0.071, +13.6\pm1.9, -21.9\pm1.1)$~km\,s$^{-1}$, 
where $U$, $V$, and $W$ are directed towards the Galactic anti-center, the 
Galactic rotation direction, and the North Galactic Pole, respectively.
These velocity components make \denis kinematically 
consistent with those of the galactic 
young disk population. Note that \citet[][]{Montes01.2} achieved a similar conclusion. 
We stress that these results are in line with the $\log{g}$ value 
derived for \denis in Sect.~\ref{subsect:xshooter_stellarparams}.

\subsection{Emission line spectrum}\label{subsect:xshooter_emissionlines}

During the X-Shooter observation \denis presented 
a weak emission line spectrum resulting from chromospheric activity.  
We detect the Balmer series up to H8. 
H$\epsilon$ is clearly resolved from \ion{Ca}{ii}\,H. 
While \ion{Ca}{ii}\,H\&K are in emission, the \ion{Ca}{ii}\,IRT is not detected, 
and no NIR emission lines are seen.  
The Na\,I doublet at $\lambda\lambda$\,588.99, 589.59\,nm 
is in absorption with central emission peaks in both lines. 
In this case, deriving the chromospheric losses requires subtraction of the
photospheric absorption profile. 
We used the ``spectral synthesis" technique, based 
on the comparison between the target spectrum and the best fit synthetic 
spectrum ($T_{\rm eff}=2500$\,K and $\log{g}=4.0$), that we use as 
``reference spectrum", to account for the non-active photospheric emission. 
The difference between the observed and the reference spectrum provides, 
as residual, the net chromospheric line emission, which can be integrated 
to find the total radiative losses in the line 
\citep[see, e.g.,][]{Frasca94.1}. 
%
\begin{figure*}[t]
\begin{center}
\includegraphics[width=15cm]{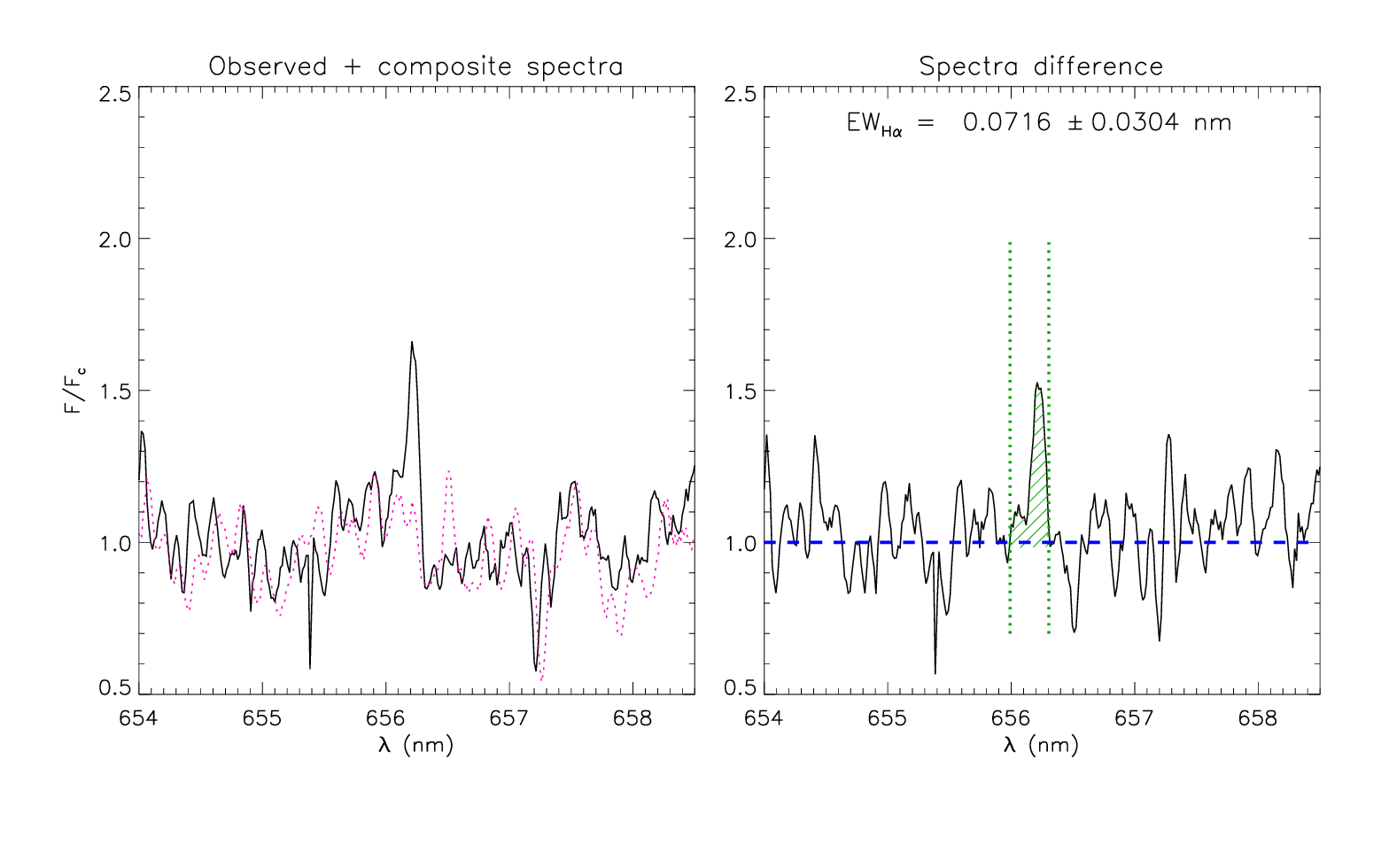}
\includegraphics[width=15cm]{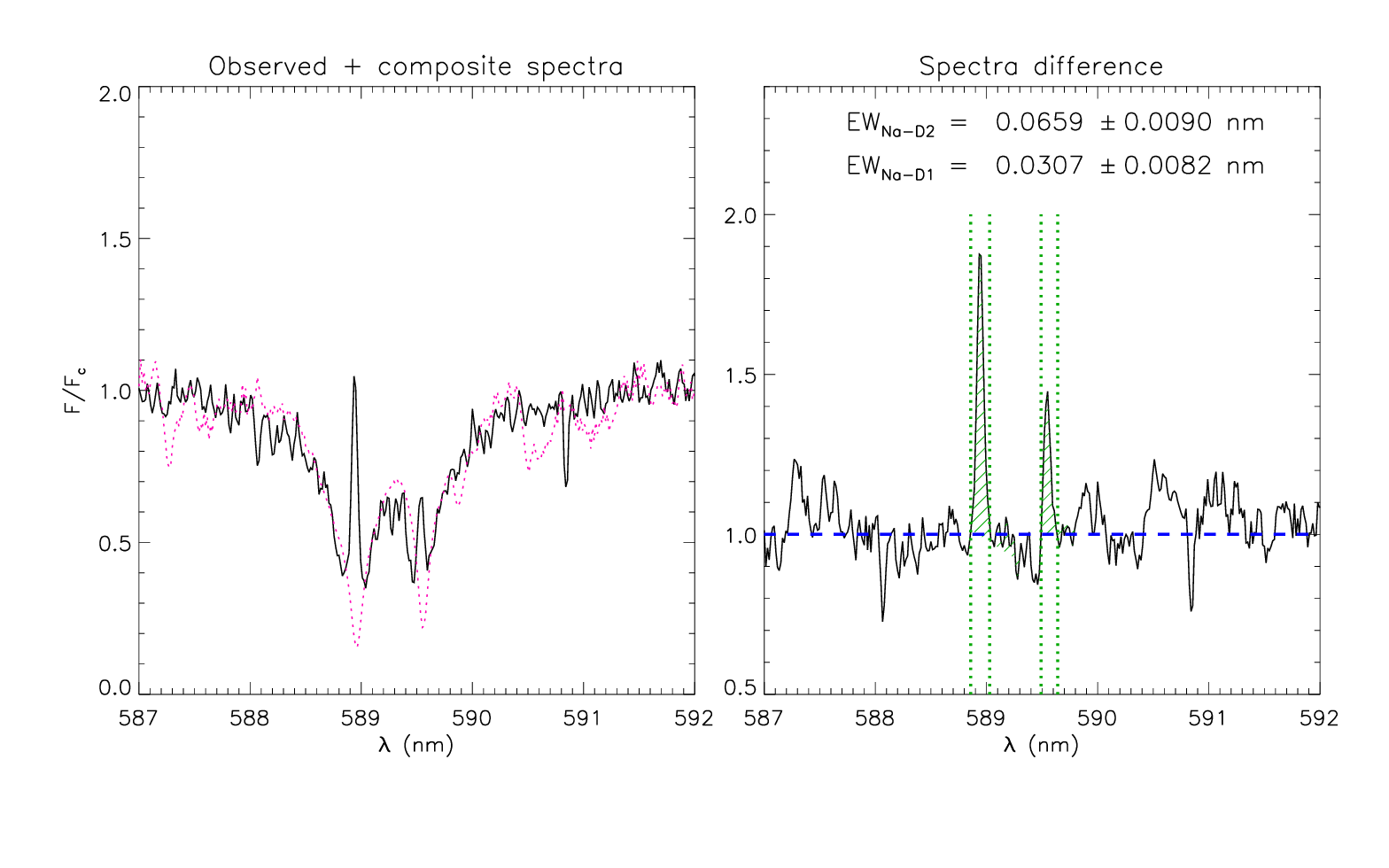}
\caption{{\it Left panels:} Observed, continuum-normalized X-Shooter spectrum 
of \denis (solid line) in the H$\alpha$ and \ion{Na}{i}\,D regions 
({\em top and bottom}, respectively) together with the 
synthetic template (dotted line). {\it Right panels:} Difference of observed 
and template spectra. The hatched areas represent the excess emissions that 
have been integrated to get the net equivalent width.}
\label{fig:Halpha_Na}
\end{center}
\end{figure*}

We applied this technique to the H$\alpha$ and the Na\,I\,D lines while 
for the other emission lines it was not necessary because the photospheric
contribution in absorption is negligible. 
In Fig.~\ref{fig:Halpha_Na} left 
we show the spectrum of \denis in 
the H$\alpha$ and \ion{Na}{i} regions, together with the reference synthetic 
spectrum, which 
is representative for the star in absence of chromospheric activity. 
The H$\alpha$ profile is clearly in emission, while the \ion{Na}{i} lines are 
in absorption with an emission core. The residual equivalent width 
of the lines has been measured by integrating the full emission profile in 
the subtracted spectrum (see right panel of Fig.~\ref{fig:Halpha_Na}). 
The error on the 
$EW$ ($\sigma_{EW}$) was evaluated by multiplying the integration range by 
the photometric error on each point. The latter was estimated by the standard 
deviation of the observed fluxes on the difference spectra in two spectral 
regions near the line. Final line residual equivalent widths and line fluxes 
%
for Na\,D and H$\alpha$ are given in Table~\ref{tab:lines} together
with those of all other emission lines. 
Note, that no helium lines are detected (e.g. ${\rm He\,I\,D_3}$\,587.6\,nm) although
these lines are often detected in the spectra of flare stars. 
For all lines in Table~\ref{tab:lines} except for Na\,D and H$\alpha$  
we have obtained the line flux by directly integrating the line profile 
in the observed spectrum. The uncertainties are determined from the local
fluctuation of the continuum. 
\begin{table}\begin{center}
\caption{Emission line parameters for \denis}
\label{tab:lines}
\begin{tabular}{lccc} \hline
Line  & $\lambda_{\rm line}$ & $EW_{\rm line}$ & $\log{f_{\rm line}}$ \\
      & [nm]      & [nm]      & [${\rm erg/cm^2/s}$] \\ \hline
                Ca K & $   393.356$ & $  -1.76 ^{  -4.74}_{+   0.98}$ & $  -15.72 ^{+ 0.03}_{- 0.02}$ \\
                Ca H & $   396.833$ & $  -1.19 ^{  -6.67}_{+   0.76}$ & $  -15.86 ^{+ 0.03}_{- 0.05}$ \\
                  H8 & $   388.893$ & $  -0.12 ^{  -0.17}_{+   0.06}$ & $  -16.71 ^{+ 0.06}_{- 0.11}$ \\
         H$\epsilon$ & $   396.995$ & $  -0.19 ^{  -0.44}_{+   0.11}$ & $  -16.56 ^{+ 0.14}_{- 0.14}$ \\
           H$\delta$ & $   410.159$ & $  -0.19 ^{  -0.34}_{+   0.09}$ & $  -16.49 ^{+ 0.06}_{- 0.09}$ \\
           H$\gamma$ & $   434.035$ & $  -0.19 ^{  -0.13}_{+   0.07}$ & $  -16.13 ^{+ 0.07}_{- 0.09}$ \\
            H$\beta$ & $   486.110$ & $  -0.13 ^{  -0.02}_{+   0.02}$ & $  -15.80 ^{+ 0.03}_{- 0.05}$ \\
           Na\,I\,D1 & $   589.577$ & $ -0.07^{-0.01}_{+0.01}$ & $ -15.94^{+0.10}_{-0.14}$  \\
           Na\,I\,D2 & $   588.980$ & $ -0.03^{-0.01}_{+0.01}$ & $ -15.61^{+0.05}_{-0.07}$  \\
           H$\alpha$ & $   656.240$ & $ -0.07^{-0.03}_{+0.03}$ & $ -14.94^{+0.14}_{-0.25}$  \\
\hline
\end{tabular}
\end{center}\end{table}

A parameter of particular relevance for this study of multi-wavelength activity
is \loglhalbol. 
Taking account of the distance, the observed H$\alpha$ flux corresponds to 
$L_{\rm H\alpha}= (3.8 \pm 0.8) \cdot 10^{24}$\,erg~s$^{-1}$.
An alternative means of computing line luminosities involves the surface flux and
radius of the star.
To derive the luminosity of a given line from the equivalent width, the surface flux
of the continuum ($F_{\lambda}^{C}$) must be known. 
We estimated the stellar radius from the COND03 models \citep[][]{Baraffe03.1} 
that best match the temperature and gravity, while $F_{\lambda}^{C}$ was estimated 
from the flux calibrated synthetic spectrum of the same $T_{\rm eff}$ and 
$\log{g}$.
In this way, we find $R_{*}=0.13$\,R$_{\odot}$ and 
$F_{\rm H\alpha}^{C} \sim 8.5 \cdot 10^4\,{\rm erg/cm^2/s/nm}$.  
The resulting line luminosity is 
$L_{\rm H\alpha}= (6.3 \pm 2.6) \cdot 10^{24}$\,erg/s, which is in agreement within the
uncertainties with the value derived from the integrated line flux.    
We compute the $L_{\rm H\alpha}/L_{\rm bol}$ value from the mean of the H$\alpha$
luminosities computed with the two methods, and obtain $\loglhalbol = -5.3$.  
This is similar to the value presented by 
\cite{Schmidt07.1} ($\loglhalbol = -5.2$) but much lower than 
the measurement of \cite{Fuhrmeister04.1} during a large flare ($\loglhalbol = -4.0$).

\section{Multi-wavelength picture of activity on UCDs}\label{sect:multi_lambda}

To better understand the nature of \denis we now 
put its multi-wavelength properties into context with earlier M dwarfs and 
with other UCDs. For this purpose we have compiled a list with activity
diagnostics for UCDs comprising X-ray, H$\alpha$ and radio luminosities, rotational
velocity, and magnetic field strength.

\subsection{Flux-flux relations}\label{subsect:flux_flux}

The relative radiative losses in different parts of the chromosphere and the corona 
can be quantified by relations between the radiation emitted in different lines. 
The connection between various chromospheric lines and soft X-ray emission 
has been shown for various samples of
late-type main-sequence stars to result 
in power law relations between the fluxes, i.e. linear
relations in log-log form of the type $\log{F_1} = c_1 + c_2 \cdot \log{F_2}$,
where $c_2$ is the power law slope and $F_1$ and $F_2$ are the surface fluxes
of the two diagnostics. 
Observational results and also their interpretation have been ambiguous if
not contradictory. In early studies, 
a spectral type dependence of the slopes has repeatedly been found in comparisons
of X-ray flux to chromospheric fluxes (H$\alpha$, Ca\,II\,H+K) with M dwarfs having
a shallower slope than G dwarfs. 
These differences have been attributed either to an excess of chromospheric emission
in M dwarfs \citep[e.g.][]{Schrijver87.1, Rutten89.1} 
or to a deficiency of coronal emission in G dwarfs \citep{Doyle89.1}. 
However, the lack of simultaneousness in the observations 
and the use of different techniques to determine chromospheric 
fluxes by distinct authors make this hypothesis frail. 
\cite[][henceforth referred to as MA11]{Martinez-Arnaiz11.0} 
found that all the main-sequence dwarf stars seem to follow the same relation between 
the surface flux in X-rays and H$\alpha$, while differences are found for 
very active K and M dwarfs when comparing X-rays with Ca~II lines.

\cite{Schrijver89.1} ascribe the minimum flux level observed for a given spectral 
type in chromospheric and transition region emissions 
to a `basal' chromospheric 
flux that is due to non-magnetic heating of the outer atmosphere and 
needs to be subtracted when investigating flux-flux relations to probe magnetic activity. 
In fact, only when the `basal' chromospheric 
flux is subtracted, clear relations are found. We refer the 
interested reader to Sections 3.1 and 5.1 of MA11
for a complete description of techniques, previous works 
and results on this issue. 

MA11 have presented flux-flux relations combining a sample of nearly $300$ 
single dwarf stars with spectral types from F to mid-M.
Given the above caveats, it seems in order to examine 
flux-flux relations for stars of different spectral type separately. 
We examine here for the first time chromospheric and coronal 
flux-flux relations for UCDs by comparing 
\denis to the subsample of M dwarfs from MA11. The sample of MA11 
comprised $39$ single dwarf stars
with spectral types from M0 to M5. MA11 identified two groups of M dwarfs. 
The `inactive' M dwarfs follow the same relationships between H$\alpha$ and
various calcium lines (\ion{Ca}{ii}\,H\&K, \ion{Ca}{ii}\,IRT) than those derived for
the stars with earlier spectral type in the MA11 sample. 
The `active' M dwarfs  
present an excess of H$\alpha$ flux with respect to these relations. The `active'
and `inactive' groups are
distinguished by the former ones being saturated in terms of $L_{\rm X}/L_{\rm bol}$
and being younger than the average. 

MA11 have derived power law relations between calcium lines, H$\alpha$ and
$L_{\rm x}$ for their total stellar sample.  
Here, we take into consideration only their M dwarfs and extend the study to the
remaining Balmer lines. 
Equivalent widths for H$\alpha$ to H$\epsilon$ have been measured from the spectra
presented by MA11 \citep[see][for a detailed description of the process]{LopezSantiago10.1}
but have not been published so far. 
To obtain the surface fluxes for the MA11 sample we 
make use of the $\chi$ factors given by \cite{West08.0}. The $\chi$ factor is defined
as $\chi_{\rm i} = f_{\rm cont, i}/f_{\rm bol}$ where 
$f_{\rm cont, i}$ is the continuum flux near a certain emission line $i$ 
and $f_{\rm bol}$ the bolometric flux. \cite{West08.0} have calibrated $\chi_{\rm i}$
as a function of spectral subclass from M0 to L0 from a sample of nearby dwarfs 
with flux calibrated spectra. 
Multiplying $\chi_{\rm i}$ with the observed line equivalent width yields 
the fractional line
flux, $f_{\rm line, i}/f_{\rm bol}$, which is equivalent to the fractional line 
luminosity, $L_{\rm line, i}/L_{\rm bol}$. We obtain the individual 
bolometric luminosity and the stellar radii required to convert luminosity 
to surface flux from the models by \cite{Siess00.1} assuming that all stars 
from the MA11 sample are on the zero-age main-sequence. 

As a result of the low spectral resolution of the SDSS spectra, 
\cite{West08.0} give a joint $\chi$ value for \ion{Ca}{ii}\,H+K and H$\epsilon$, 
while these three lines are
resolved in the M dwarf sample from MA11.
Given the negligible variation of the continuum flux in the small wavelength range
comprising these three lines we can use the same $\chi$ factor to derive individual line
fluxes for \ion{Ca}{ii}\,H, \ion{Ca}{ii}\,K, and H$\epsilon$. 

\begin{figure*}[t]
\begin{center}
\parbox{18cm}{
\parbox{9cm}{
\includegraphics[width=8cm]{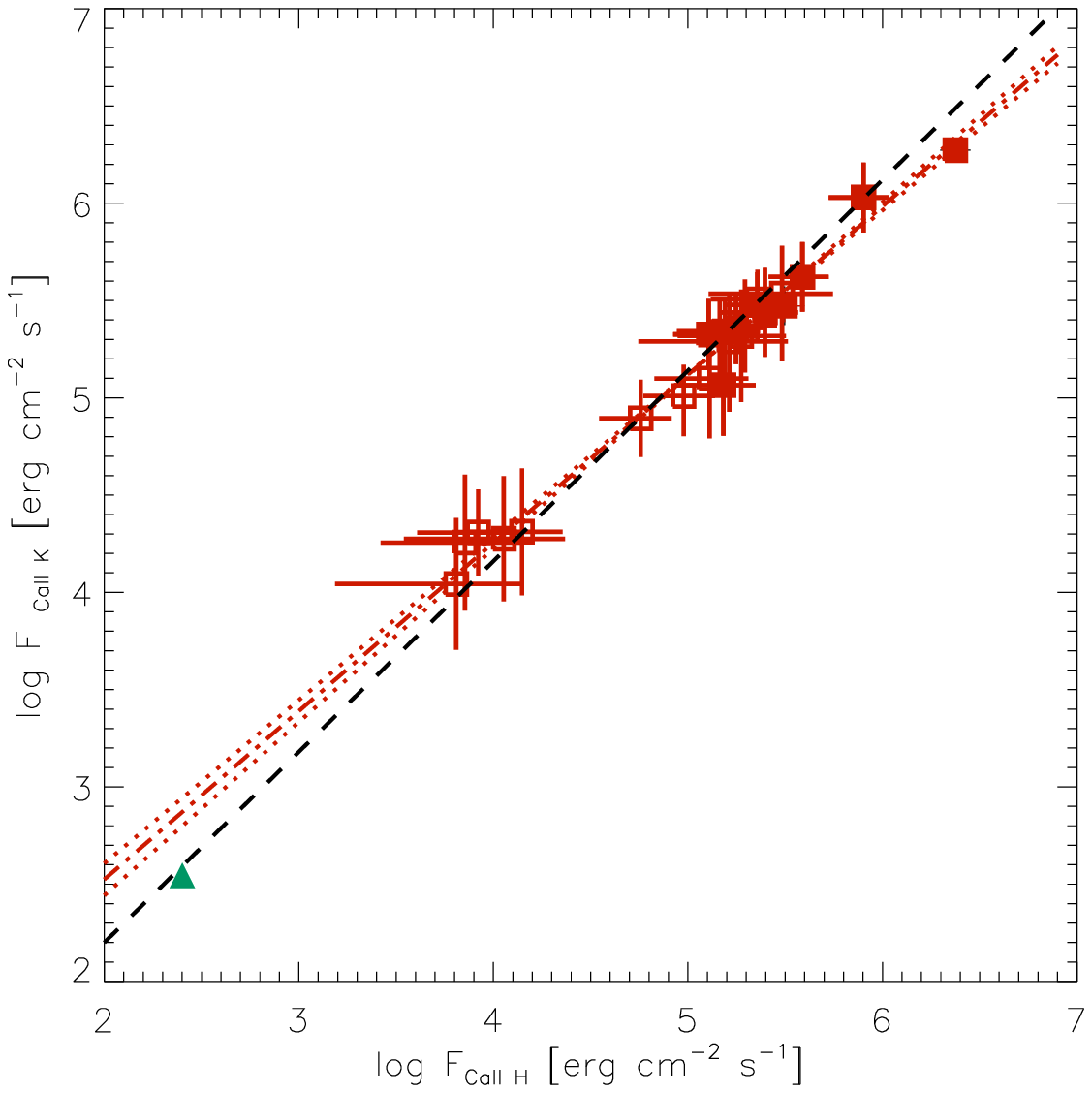}
}
\parbox{9cm}{
\includegraphics[width=8cm]{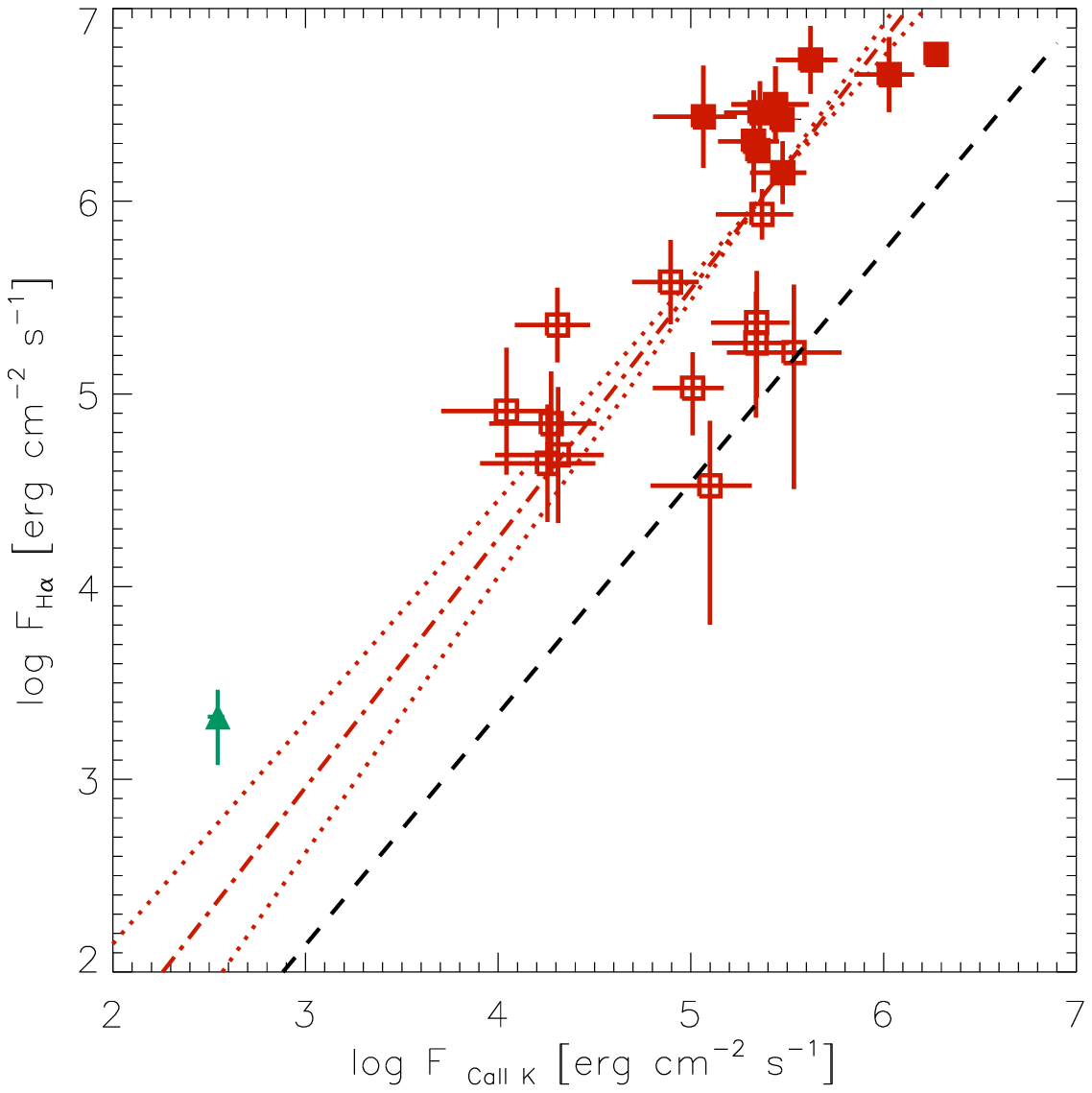}
}
}
\parbox{18cm}{
\parbox{9cm}{
\includegraphics[width=8cm]{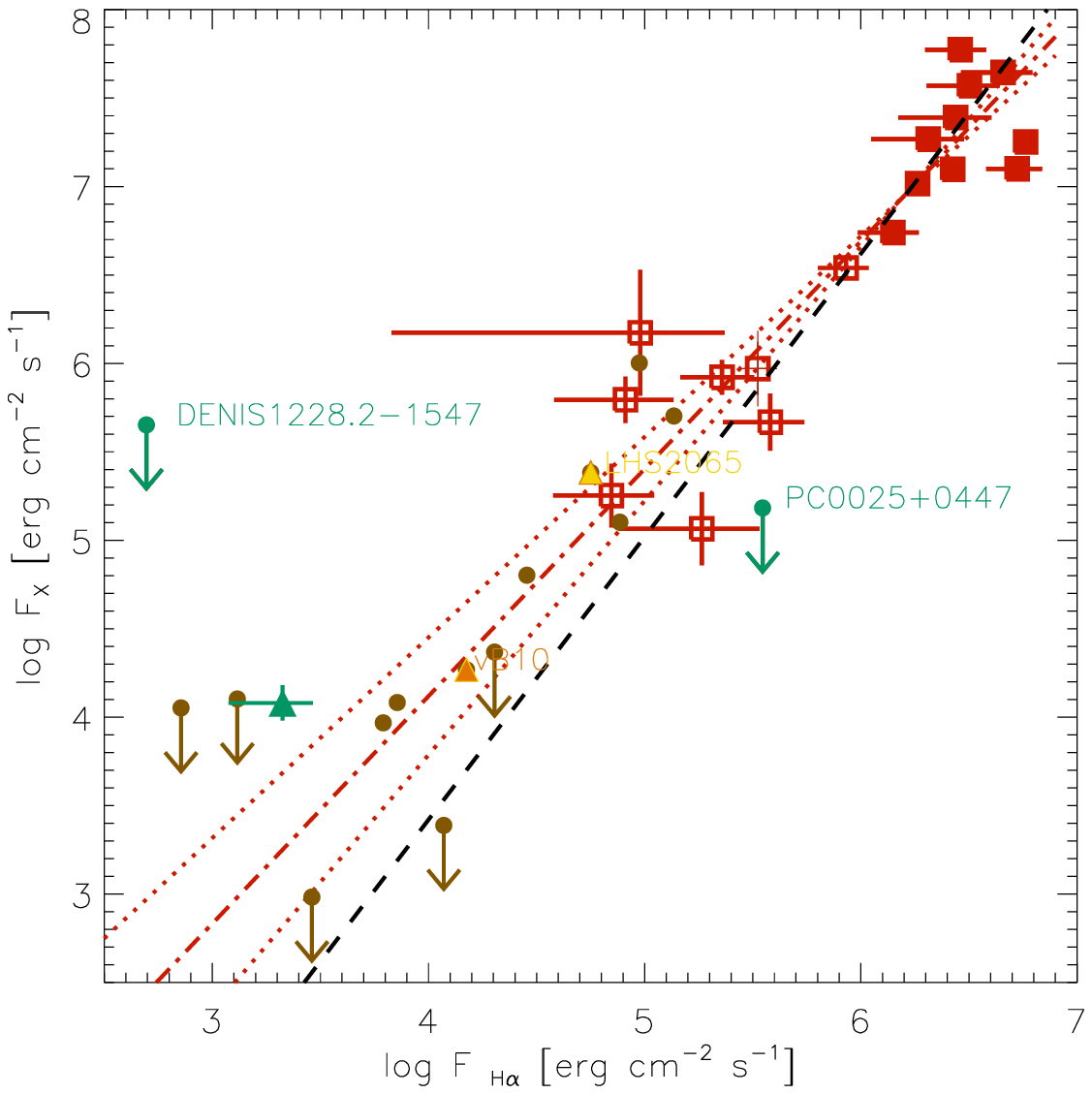}
}
\parbox{9cm}{
\includegraphics[width=8cm]{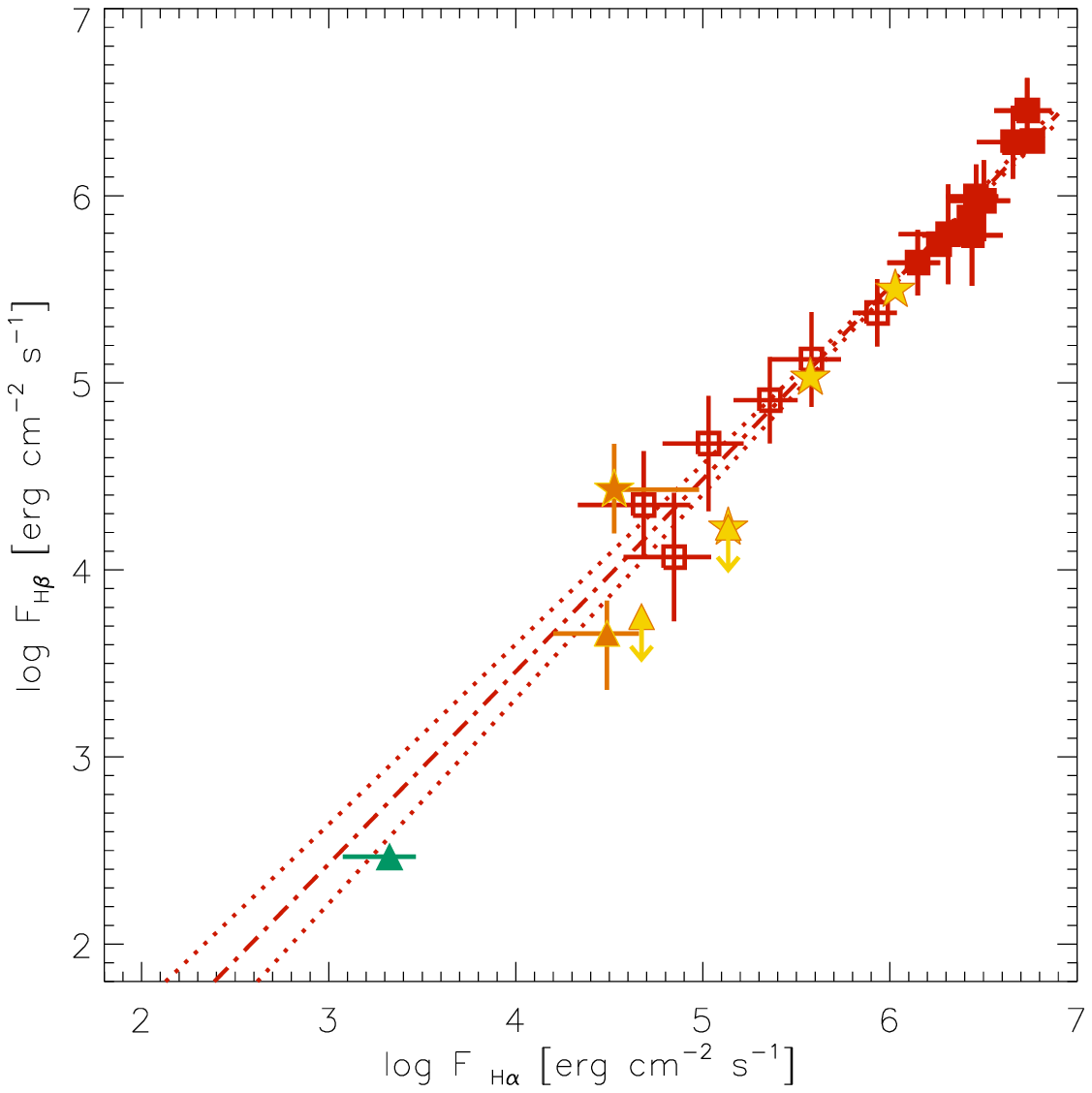}
}
}
\caption{Relations between chromospheric and coronal activity diagnostics
for the sample of M dwarfs from MA11 (red) with linear regression fits
(red dash-dotted) and their variance (red dotted)
and for comparison the fits derived by \cite{MartinezArnaiz11.1} for their whole sample
of F-M dwarfs (black dashed):  
(a) \ion{Ca}{ii}\,K vs. \ion{Ca}{ii}\,H flux, 
(b) H$\alpha$ vs. \ion{Ca}{ii}\,K flux, 
(c) X-ray vs. H$\alpha$ flux and 
(d) H$\beta$ vs. H$\alpha$ flux.  
Also plotted, but not included in the fits, is 
in all panels \denis (green triangle), 
in panel\,(c) 
UCDs from the literature (brown circles, orange and yellow triangles) 
and two UCDs for which we present new X-ray data 
(\pcstar and \denisstar; green circles), 
and in panel\,(d)
two UCDs with both H$\alpha$ and H$\beta$ measurements from the literature
(vB\,10 in orange and LHS\,2065 in yellow; triangles represent quiescence and
star symbols the flare state; the data points 
for LHS\,2065 correspond to four consecutive spectra tracking the decay of a large flare
and show that in the course of the flare evolution its 
H$\alpha$ and H$\beta$ fluxes vary along the flux-flux relationship for earlier M dwarfs).}
\label{fig:flux_flux}
\end{center}
\end{figure*}

In Fig.~\ref{fig:flux_flux} we reproduce some of the plots from MA11
including only their M stars. We examine only those relations  
for which we have measured data for \denis, i.e. the study of the \ion{Ca}{ii}\,IRT is
excluded. 
The flux for the \ion{Ca}{ii}\,IRT line at $\lambda 8498$\,\AA~ 
predicted for \denis from the
flux-flux relation of \ion{Ca}{ii}\,K versus \ion{Ca}{ii}\,IRT\,$\lambda$8498 from MA\,11 
is $1.7 \cdot 10^{-16}\,{\rm erg/cm^2/s}$. This is within the noise level in the
X-Shooter spectrum making plausible its non-detection. 
Note that MA11 have derived surface fluxes from empirical relationships between 
continuum flux at the line position and $B-V$ color following \cite{Hall96.0}. 
Therefore, there are slight 
differences between their fluxes and those that we have derived with the $\chi$-factors. 
However, MA11 have shown in their Fig.~2 that there is a very good correlation between the
results of both methods. 
For \denis we prefer the fluxes obtained directly from the flux-calibrated spectrum
with respect to those calculated from $\chi$-factors because the former ones 
have much smaller uncertainties and are available for a larger number of lines. 

The error bars in Fig.~\ref{fig:flux_flux} reflect 
uncertainties in the measurement of the line equivalent width 
and in the $\chi$-factors, and are dominated by the latter. 
Additional errors in the assumption of the individual radii and distances
are not taken into account. Stars from MA11 with multiple observations are plotted at
the mean of the measured flux with an error bar that represents the standard
deviation of the individual measurements. 
The `active' group defined by MA11 is recognizable as filled squares. 

We performed linear regression fits to this subsample of M dwarfs  
(dash-dotted lines in Fig.~\ref{fig:flux_flux}) 
to be compared to the linear regressions derived by MA11 
and corrected by \cite{MartinezArnaiz11.1} 
for the whole sample of F-M dwarfs (dashed lines). 
In Table~\ref{tab:Hlines} we give the values obtained for our linear regressions,
including Balmer lines that were not included in MA11 and for which data has been measured
but not published before. 
Note, that we have carried out the fitting process using
different techniques described by \citet{Isobe90.1}. 
We report in Fig.~\ref{fig:flux_flux}  
and in Table~\ref{tab:Hlines} the results from the least-squares bisector
regression, a method that is adequate in cases where the intrinsic scatter of the data dominates
over the measurement errors and that treats the X and Y variables symmetrically.
The same technique was used by MA11 and \cite{MartinezArnaiz11.1},  
making the results directly comparable.  
Our power-law slope of $F_{\rm Ca\,K}$ vs. $F_{\rm Ca\,H}$ 
agrees well with that derived by \cite{MartinezArnaiz11.1}, while 
that of $F_{\rm H\alpha}$ vs. $F_{\rm Ca\,K}$ 
and of $F_{\rm x}$ vs. $F_{\rm H\alpha}$ are slightly different from 
the results of \cite{MartinezArnaiz11.1} considering the $1\,\sigma$ uncertainties. 
\denis is shown as green triangle in all plots and is not included in the fitting process
(nor is any of the other UCDs that are shown in some of the plots as described below). 
%

\begin{table}
\begin{center}
\caption{Linear fit coefficients of flux-flux relationships for the M {\bf dwarfs} from MA11}
\label{tab:Hlines}
\begin{tabular}{ccrrr} \\ \hline
Line\,1 &   Line\,2 &    $ N_*$ &  \multicolumn{1}{c}{$c_1$}  &  \multicolumn{1}{c}{$c_2$} \\
\hline
     Ca\,K    & Ca\,H  &   24 & $ 0.82 \pm  0.46$ & $ 0.86 \pm  0.08$ \\
\hline
     H$\alpha$ & Ca\,K  &  22 & $-1.32 \pm  0.85$ & $ 1.26 \pm  0.15$ \\
\hline
    X-rays  & H$\alpha$  & 18 & $ -1.02 \pm  0.94$ & $ 1.29 \pm  0.15$ \\
\hline
 H$\alpha$ & H$\beta$  & 16 & $-0.66 \pm  0.41$ & $ 1.03 \pm  0.07$ \\
\hline
 H$\alpha$ & H$\gamma$  &  14 & $-1.33 \pm  0.32$ & $ 1.06 \pm  0.05$ \\
\hline
 H$\alpha$ & H$\delta$  &  13 & $-2.36 \pm  0.47$ & $ 1.20 \pm  0.08$ \\
\hline
 H$\alpha$ & H$\epsilon$  & 10 & $-14.66 \pm 5.63$ & $2.96 \pm 0.87$ \\
 \hline
  H$\beta$ & H$\gamma$  &  14 & $-0.94 \pm  0.18$ & $ 1.08 \pm  0.03$ \\
\hline
  H$\beta$ & H$\delta$  &  13 & $-1.40 \pm  0.14$ & $ 1.14 \pm  0.03$ \\
\hline
\end{tabular}
\tablefoot{Linear regression fits are of the type $\log{F_1} = c_1 + c_2 \cdot \log{F_2}$ and
col.3 gives the number of stars in the sample.}
\end{center}
\end{table}

First, we concentrate on some features in the derived flux-flux relations. 
While the \ion{Ca}{ii}\,H and \ion{Ca}{ii}\,K lines show a tight correlation, the plot of 
H$\alpha$ versus \ion{Ca}{ii}\,K flux displays a very large scatter. Rather than a clear relation 
there seem to be two separate clouds of data for the `active' and the 
`inactive' stars. We note, that the scatter in this plot becomes even larger
if stars of spectral types F-K are included. 
This is also the reason for the vertical offset of our linear regression with respect to
the results of \cite{MartinezArnaiz11.1} in Fig.~\ref{fig:flux_flux}(b). 
In fact, MA11 defined their `active' and `inactive' sample 
on the basis of this plot.  
X-ray versus H$\alpha$ flux shows a much clearer correlation, although also this
plot shows considerable scatter. 
As an example for the flux-flux relations between Balmer lines, that are presented here
for the first time for the MA11 sample, 
in Fig.~\ref{fig:flux_flux}\,(d) H$\beta$ vs. H$\alpha$ is displayed.  
Very tight correlations are observed for all combinations of Balmer line fluxes 
(see Table~\ref{tab:Hlines} for the results of the regression analysis). 

In Fig.~\ref{fig:flux_flux}(c) and~(d) 
we include data on UCDs compiled from the literature. 
To obtain the individual surface fluxes for the lines from the 
published equivalent widths we have assumed radii between
$0.12$ and $0.16\,{\rm R_\odot}$ depending on the spectral type, in accordance with
the predictions of the \cite{Baraffe03.1} model atmospheres for high-gravity ($\log{g}=4.5$) 
M7-L0 dwarfs.  
Where more than one epoch of X-ray or H$\alpha$ data is available for these
objects we adopt the minimum measured value as representative of the 
quiescent emission. 
This is justified because the spectra of M dwarfs from MA11 from which the
flux-flux relations are derived represent snapshots for individual stars and
this sample is, therefore, unlikely to be dominated by flares.\footnote{The range of
observed H$\alpha$ fluxes for the individual objects in our UCD sample is between
$0.5$ to $2$\,dex. This is larger than the typical short-term H$\alpha$ variations 
(of a factor two in $\Delta t \leq 1$\,hr) 
observed on M dwarfs \cite[e.g.][]{CrespoChacon06.1,Lee10.0,Kruse10.0}. These differences are probably due 
to the much shorter timescales examined in these latter studies and due to 
observational biases in our sample in favor of UCDs that have caught attention due to 
flaring activity either in H$\alpha$ or X-rays.
Similarly, the majority of UCDs with an X-ray detection have displayed an X-ray flare, 
and, therefore, the mean of the \underline{observed} fluxes is biased towards 
high values making the lowest observed flux a better measure for the typical emission
level. 
} 
Analogously, the position of \denis in Fig.~\ref{fig:flux_flux} 
represents its lowest observed coronal 
and chromospheric flux, i.e. the calcium and Balmer line fluxes 
from our X-Shooter spectrum and our new {\em XMM-Newton} detection.  

Compared to the extrapolation of the flux-flux relations to low fluxes, 
\denis displays several deviations: In the relation between H$\alpha$ and \ion{Ca}{ii}\,K flux
\denis has an excess of H$\alpha$ for its \ion{Ca}{ii} emission
and in terms of \ion{Ca}{ii}\,K vs. \ion{Ca}{ii}\,H it is also outside by $>3\,\sigma$
of the very tight relation. 
In X-ray vs. H$\alpha$ flux \denis is marginally compatible with the 
relation derived for the MA11 sample. 
Similarly to \denis, 
most of the other UCDs agree well with the extension of the X-ray/H$\alpha$ 
relation of early-M dwarfs to lower fluxes. 
We mention in passing that the analysis of the UCDs did not comprise the 
spectral subtraction technique and, in principle, all line fluxes may be underestimated.
However, from our experience with \denis we do not expect this to be a notable effect. 
Two UCDs (\pcstar and \denisstar) 
for which we present new X-ray data in the appendix are highlighted by green 
circles. The deviation of \pcstar from the X-ray/H$\alpha$ correlation 
is due to its notoriously bright H$\alpha$ emission 
\citep{Schneider91.1}. 
Two UCDs (LHS\,2065 and vB\,10) for which we discuss below the Balmer
decrements are also highlighted in different colors.
Further UCDs with detected H$\alpha$ emission beyond the lower end of the flux range
shown in Fig.~\ref{fig:flux_flux}\,(c) have meaninglessly high upper limits in X-rays.

\subsection{\ion{H}{i} line ratios}\label{subsect:balmer_decr}

Ratios between line fluxes are often used for flare stars to characterize the 
emitting plasma. 
We compare the Balmer decrements of \denis to those measured for other M dwarfs 
and the Sun (Sect.~\ref{subsubsect:balmer_decr_obs}) 
and to the theoretical values predicted for different assumptions for the physical conditions of the
emitting gas (Sect.~\ref{subsubsect:balmer_decr_model}).

\subsubsection{Observed Balmer decrements}\label{subsubsect:balmer_decr_obs}

The Balmer decrements are here referred with respect to H$\beta$. 
We computed the available Balmer flux ratios for the M dwarf sample from MA\,11 and for \denis. 
In addition, we computed fluxes for H$\alpha$ and H$\beta$ emission of two UCDs 
from the equivalent widths given in the literature, the M8 dwarf vB\,10 \citep{Berger08.1} 
and the M9 dwarf LHS\,2065 \citep{Martin01.1}, 
using the appropriate $\chi$ factor from \cite{West08.0}. 
For these two objects equivalent widths have been measured both during quiescence
and during flares
\footnote{For the sake of consistency 
with the literature on Balmer decrements during flares (in particular the solar data by 
\cite{JohnsKrull97.1})
we subtracted the quiescent line fluxes from the ones measured during the flare
obtaining this way the `flare-only' emission and corresponding Balmer decrement. 
In the case of LHS\,2065, where 
the flare is represented by three of the four available H$\alpha$/H$\beta$ measurements,  
we subtracted the fluxes of the last spectrum
representing the quiescence from the fluxes obtained for the three preceeding spectra.}. 
For vB\,10 and LHS\,2065 the H$\alpha$/H$\beta$ ratio is lower during the flare with respect to
the quiescent state. 
The four available measurements for LHS\,2065 representing a flare in its decaying phase follow, 
similar to the case of vB\,10, roughly the H$\beta$/H$\alpha$
flux-flux relation of earlier M dwarfs in Fig.~\ref{fig:flux_flux}. 

\begin{figure}[t]
\begin{center}
\includegraphics[width=9cm]{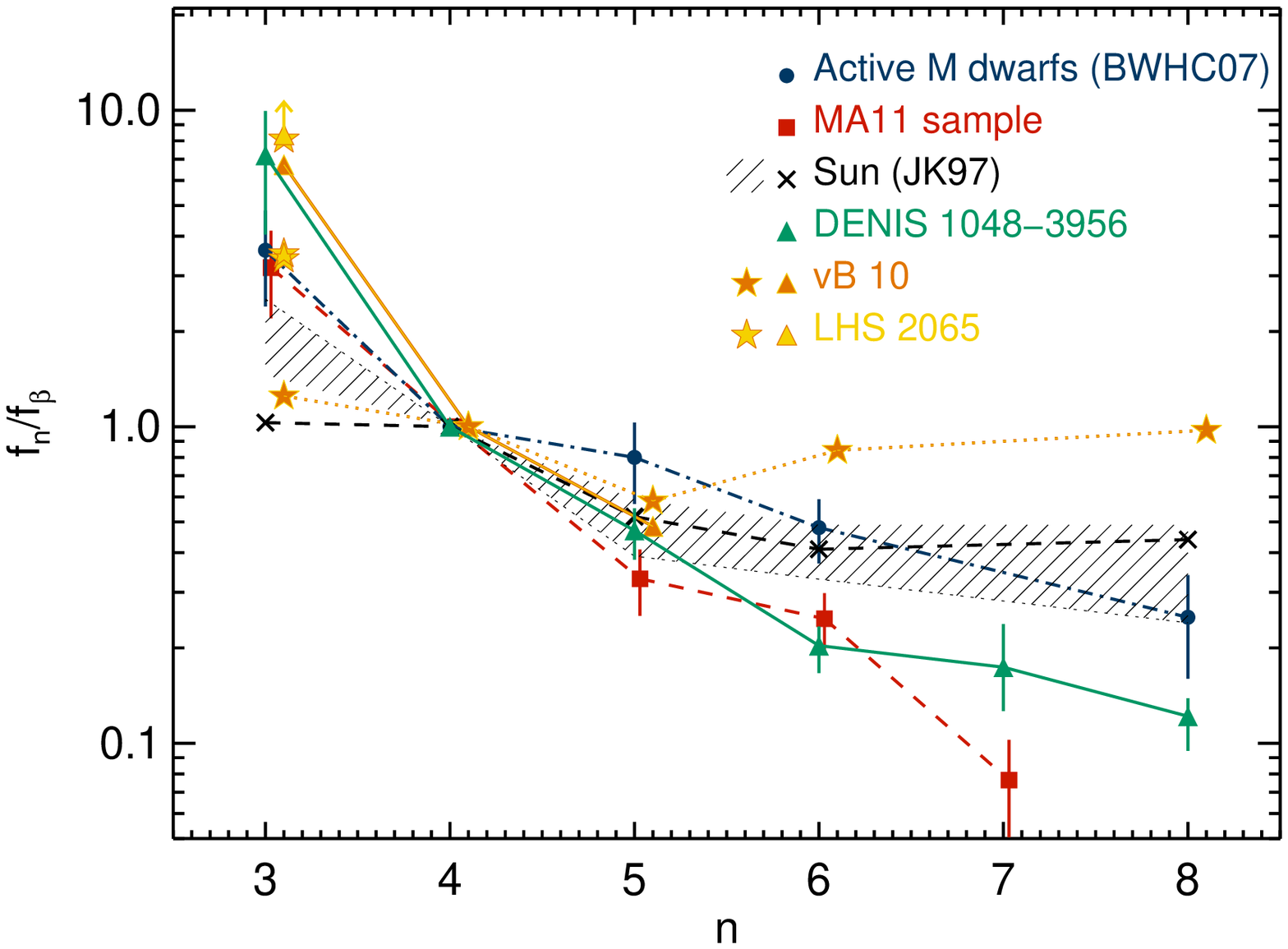}
\caption{Balmer decrements for \denis compared to the M dwarf sample from MA11, 
the active M dwarf templates from \cite{Bochanski07.1}, 
the Sun from \cite{JohnsKrull97.1},
and the UCDs vB\,10 \citep{Berger08.1} and LHS\,2065 \citep{Martin01.1}. The latter two
objects are plotted with a small horizontal offset for clarity, and 
we distinguish their flare phase (star symbols) from the quiescent state.
Similarly, for the Sun the range observed during a large flare is shown as shaded area
where the late flare phase is highlighted by a thin dotted line while the quiet Sun decrements
are connected by a dashed line.}
\label{fig:balmer_decr}
\end{center}
\end{figure}

In Fig.~\ref{fig:balmer_decr} we compare our results for \denis, vB\,10, LHS\,2065
and the early M dwarfs from MA11 to the average Balmer decrements given 
for `active' M dwarfs by \cite{Bochanski07.1} (henceforth BWHC07).  
These authors have defined template spectra for
M dwarfs from a systematic analysis of more than $4000$ 
spectra from the SDSS spectroscopic database. They define `active' stars as those
having $EW_{\rm H\alpha} > 1$\,\AA~ and H$\alpha$ emission being detected with at least 
$3\,\sigma$ above the noise level. 
If we apply the same criteria to the MA11 M dwarf sample the total number of stars
is reduced but the Balmer decrements do not change significantly. 
Fig.~\ref{fig:balmer_decr} shows also the decrements measured by \cite{JohnsKrull97.1}
(henceforth JK97) 
for the Sun during quiescence and the range observed throughout a solar flare. 
In general, M dwarfs have H$\alpha$/H$\beta$ decrement more similar to solar flares
than to the quiet state of the Sun.  
The exception is the H$\alpha$/H$\beta$ ratio. 
As compared to quiescence, the solar flare has 
higher H$\alpha$/H$\beta$ ratio, while the UCDs LHS\,2065 and vB\,10 seem to behave the 
opposite way. 
We caution that the error bars for the measurements of vB\,10, which we deliberately omit
in Fig.~\ref{fig:balmer_decr},  
are extremely large as a result of the poorly constrained $\chi$ factors for late-M dwarfs.
For LHS\,2065 no uncertainties on the equivalent widths are given in the literature. 

The higher values for the decrements of BWHC07 
with respect to those measured by us for the MA11 sample 
may be due to a spectral type dependence within the M class. 
BWHC07's decrements from H$\gamma$/H$\beta$ upward include only
stars of spectral type M4 and later, while the MA11 sample is restricted
to M0-M5 stars. 
However, extrapolating this presumible trend to UCD spectral types 
one would expect the M9 dwarf \denis to have even higher decrements than BWHC07's stars. This 
is obviously not the case. On the contrary, 
the Balmer decrements of \denis are compatible with those of the earlier-type MA11 sample.
We caution that the fluxes in the Balmer decrements have been computed in different ways
for the various samples. While for \denis we have integrated the spectrum over the wavelengths
representing the line, for both M dwarf 
samples (MA11 and BWHC07) the $\chi$ factor has been used.
BWHC07 have computed the decrements from a mean template spectrum for each spectral type,
while for the MA11 data we have evaluated each spectrum separately and then calculated the
mean of the measured decrements. 

\begin{figure}[t]
\begin{center}
\includegraphics[width=9cm]{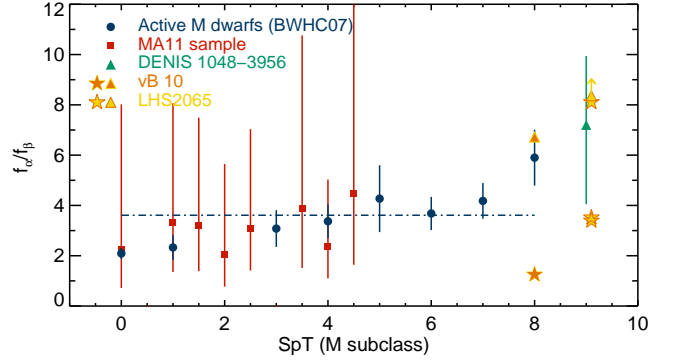}
\caption{H$\alpha$/H$\beta$ flux ratio as a function of spectral type for M dwarfs. 
Same plotting symbols as in Fig.~\ref{fig:balmer_decr}. The dash-dotted line denotes the mean over
all spectral types for the BWHC07 sample.}
\label{fig:ha_decr_spt}
\end{center}
\end{figure}

The only decrement observed for the full range of M dwarfs by BWHC07 
is H$\alpha$/H$\beta$, which is indeed in perfect agreement with the value obtained 
for the MA11 stars. BWHC07 remark that this flux ratio 
seems to become gradually stronger towards later spectral type. 
We display their measurements in Fig.~\ref{fig:ha_decr_spt}
together with our results for the MA11 sample. The latter one has very large error
bars due to the small number of objects (between $1-4$) in each spectral type bin
and the error propagation into the flux ratio of the uncertainties of the $\chi$ factors. 
The latest spectral type
for which \cite{Bochanski07.1} present Balmer decrements is M8. The value for this
spectral type derived from the SDSS sample is in perfect agreement 
with the quiescent value of vB\,10. 
Fig.~\ref{fig:ha_decr_spt} shows also that our measurement for 
\denis extends the trend at the cool end with a sharp rise of the 
H$\alpha$/H$\beta$ decrement. Although our formal error bars for \denis
are considerable the result is bolstered by the quiescent decrement of LHS\,2065,
which is actually a lower limit. 
(Recall, however, that the errors for vB\,10 and LHS\,2065, not shown in the plot, 
are even larger than those of \denis.) 

\subsubsection{Calculated Balmer decrements}\label{subsubsect:balmer_decr_model}

To determine the physical conditions in the region emitting the Balmer lines we calculated
the flux ratios for different assumptions: optically thick and optically thin gas in LTE
and Case\,B recombination.

\begin{figure*}[t]
\begin{center}
\parbox{18cm}{
\parbox{9cm}{
\includegraphics[width=9cm]{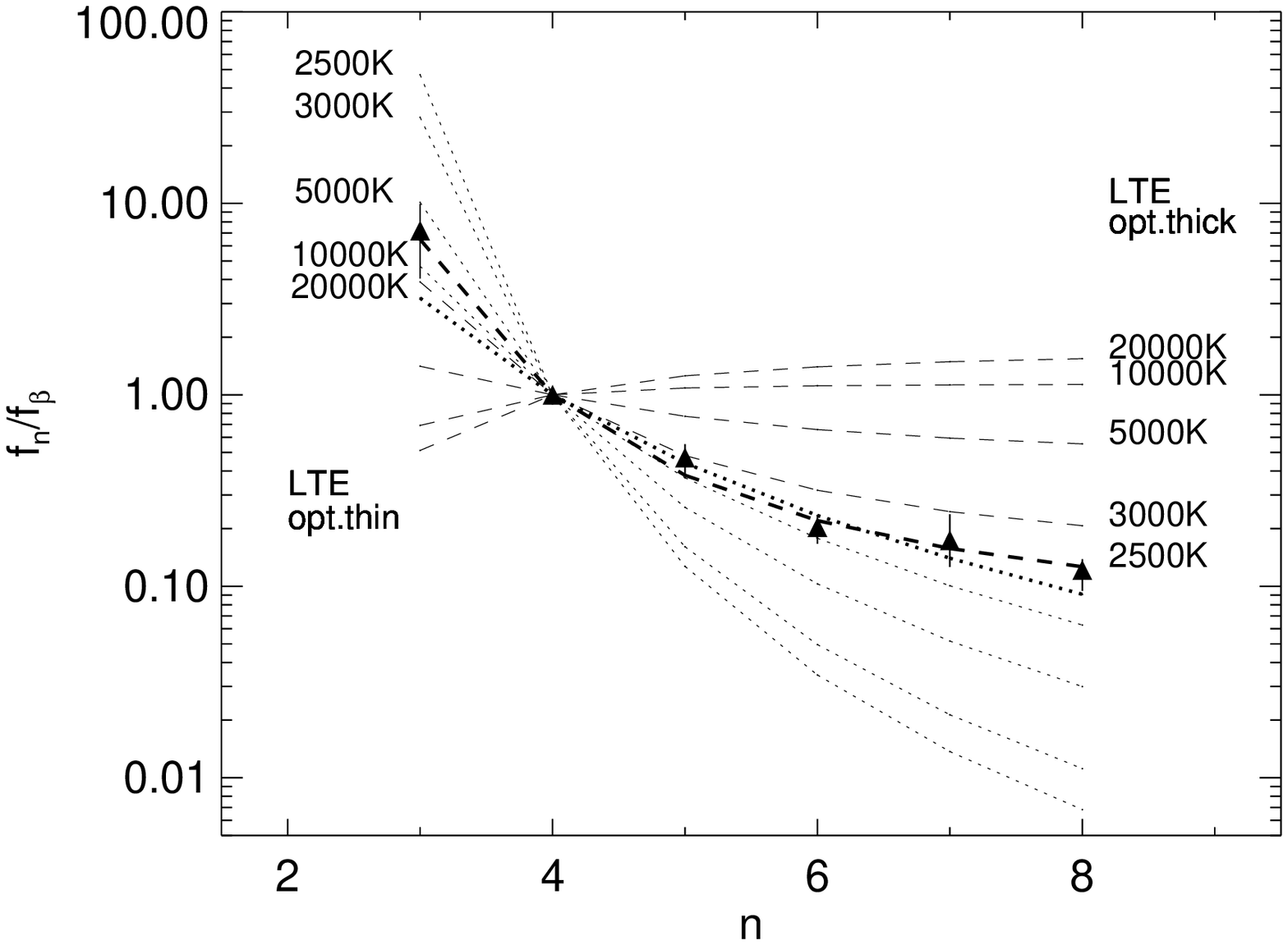}
}
\parbox{9cm}{
\includegraphics[width=9cm]{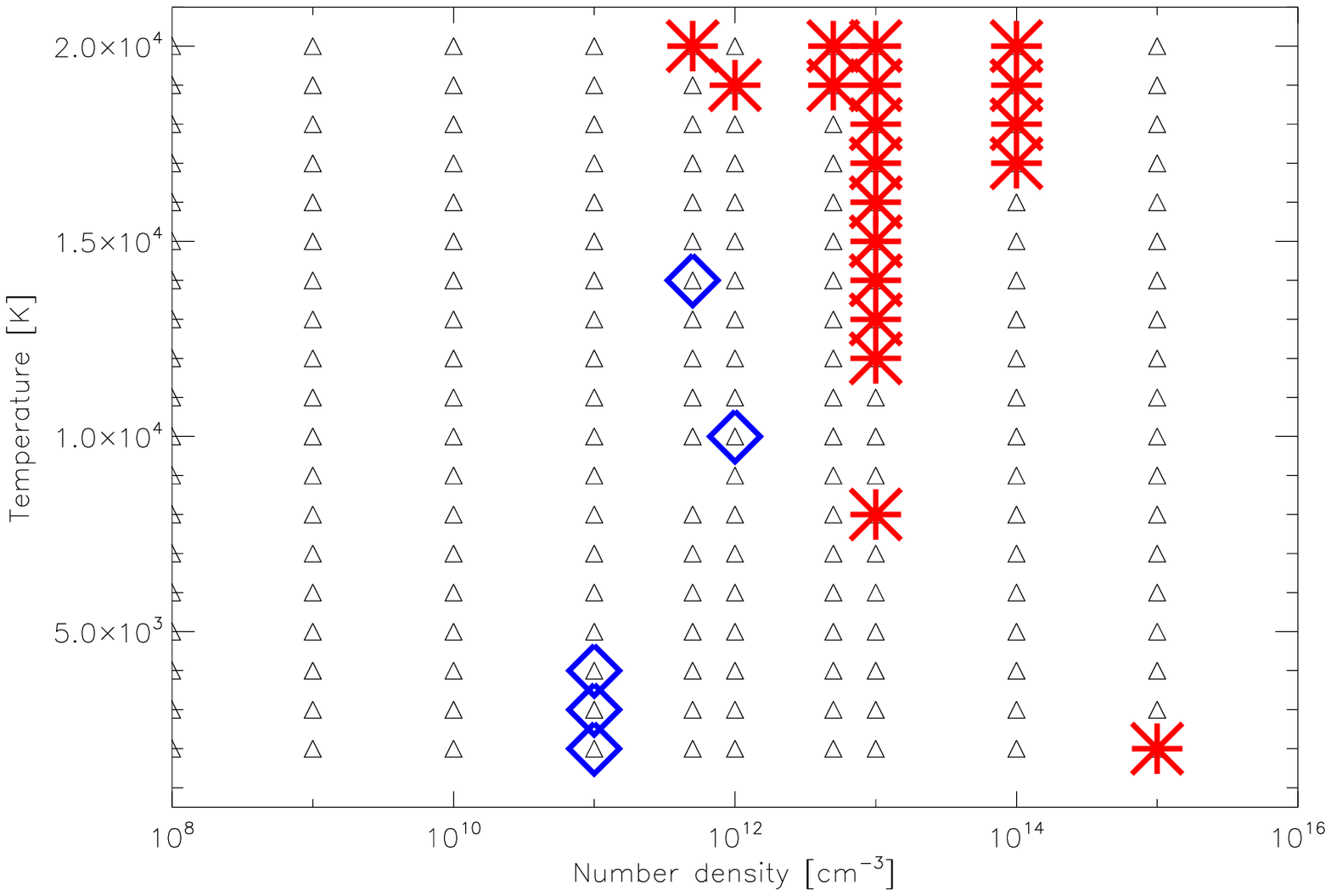}
}
}
\caption{Theoretical predictions for Balmer decrements: 
(left) -- Optically thick (dashed) and optically thin (dotted) gas in LTE and observed flux ratios for \denis. 
For both the optically thick and the optically thin case the model with the temperature that
best fits the observed decrements is highlighted as thick line. 
(right) -- Grid of temperature and electron density for a recombining hydrogen plasma. Models compatible
with the observed Balmer decrements of \denis are marked by large colored symbols: 
blue diamonds (H$\alpha$/H$\beta$) and red asterisks (the higher-n decrements).}
\label{fig:balmer_decr_models}
\end{center}
\end{figure*}

For the optically thick LTE case we computed the Balmer flux ratios according to Eq.~1 in \cite{Bary08.0}
from the Planck function. 
Optically thin LTE level populations are described by the Boltzmann distribution. We have adopted
Einstein coefficients and oscillator strengths for the Balmer lines from 
the {\em National Institute of Standard \& Technology Atomic Spectra Database}\footnote{The NIST
Atomic Spectra Database is available at http://physics.nist.gov/PhysRefData/ASD/lines\_form.html}. 
A range of temperatures from $2000$\,K to $20000$\,K is explored for both optically thin and 
thick LTE conditions. 
The predicted Balmer decrements are displayed in Fig.~\ref{fig:balmer_decr_models}\,(left) together with the
observed values for \denis. 
Optically thin LTE emission describes the trend of the Balmer decrements for \denis best
for a temperature of $20000$\,K. 
Optically thick LTE plasma fits the data only for a very low temperature 
of $2500$\,K. 
Note that this value corresponds to the temperature in the photosphere of \denis 
(see Sect.~\ref{subsect:xshooter_stellarparams}). 

The electron cascade and the population densities of the energy levels in recombining hydrogen
plasma (Case\,B) have been determined using the method of \cite{Storey95.0}. 
We performed a grid of 191 models
spanning temperatures in the range $2000$\,K to $20000$\,K and number densities of
electrons in the range $10^8$ to $10^{15}$ cm$^{-3}$. 
Models are produced and fitted to all observed Balmer decrements of \denis. 
Fig.~\ref{fig:balmer_decr_models}\,(right) shows the parameter
space, where each black triangle represents a model. We found no temperature/density 
combination that could simultaneously reproduce 
both the H$\alpha$/H$\beta$ ratio and the higher Balmer decrements 
within the range of the observational errors given in Table~\ref{tab:lines}. 
The H$\alpha$/H$\beta$ decrement is best fit by models (blue
diamonds) with lower plasma temperatures and densities compared to models that fit
the other Balmer decrements (red asterisks). If the Case B conditions apply, i.e. if the 
level populations of the plasma emitting the Balmer lines are 
determined by radiative cascades,  
H$\alpha$ emission and the other Balmer lines must be produced in different regions,
and H$\alpha$ may be formed in a region of very low temperature.

\subsection{Radio versus X-ray emission}\label{subsect:lrlx}

Over the last decade, a database of sensitive X-ray and radio observations 
of UCDs has slowly been building up. 
The relation between the emissions in these two wavebands 
has unveiled important differences in comparison to the X-ray/radio connection 
observed in low-mass GKM stars.
Detailed discussions are found, e.g., in \cite{Berger06.1} and \cite{Berger10.1}. 

\begin{figure}
\begin{center}
\includegraphics[width=9cm]{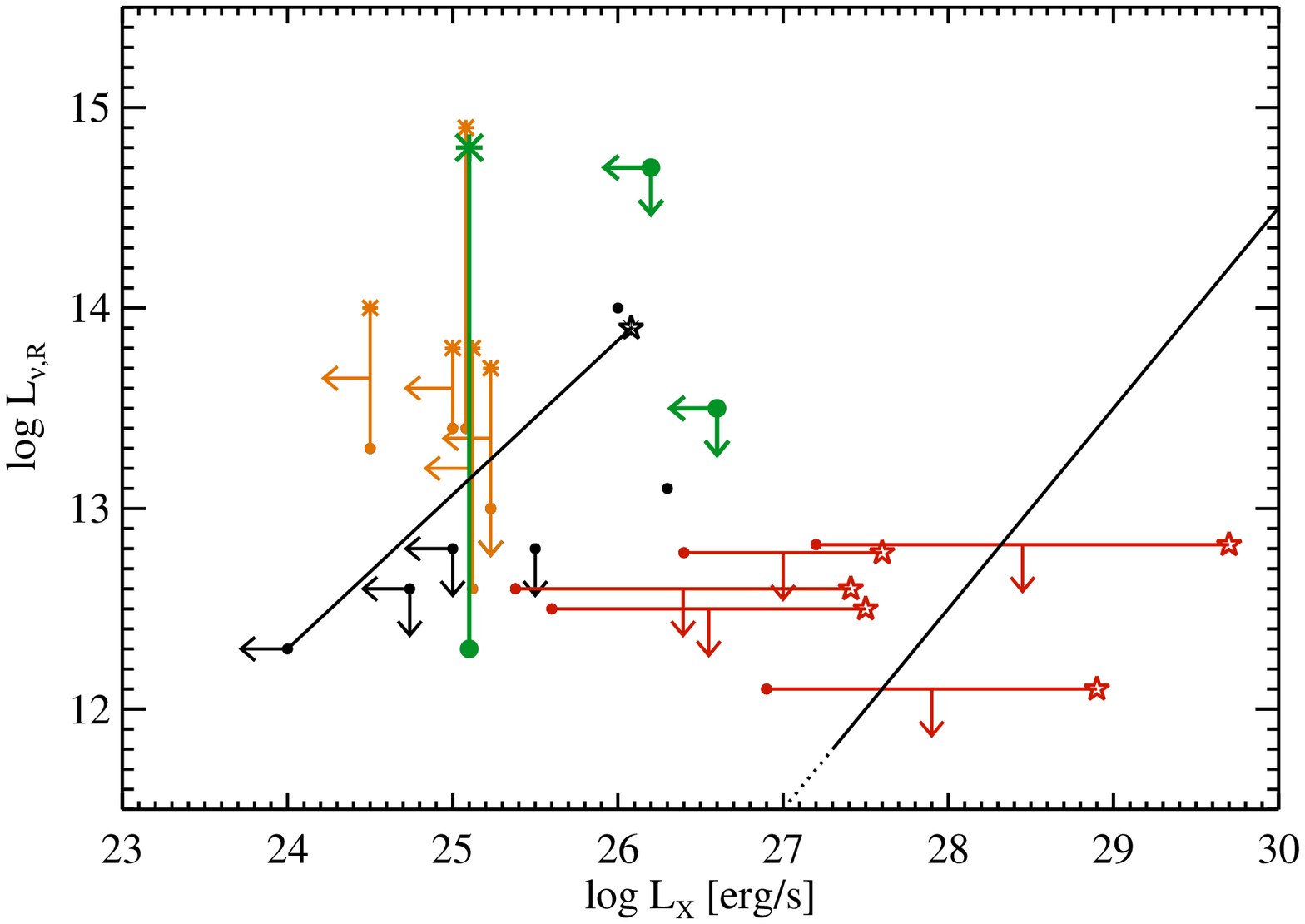}
\caption{Radio vs. X-ray luminosity for UCDs including \denis (green larger symbols connected by 
the vertical line), \pcstar and \denisstar (both also shown in green and larger). The Benz-G\"udel-relation for
more massive late-type stars is represented by the black solid line and its extrapolation to 
low luminosities is dotted. 
The lowest observed emission level of each object is shown as a circle. 
Two groups with distinct flaring behavior in the radio (orange with 
asterisks for the flares)
and X-ray (red with star symbols for the flares) bands are identified. 
Objects that can not be assigned to any of the groups (yet) are shown in black.}  
\label{fig:lr_lx}
\end{center}
\end{figure}

Making use of data collected in previous studies, 
we show in Fig.~\ref{fig:lr_lx} radio versus X-ray luminosity for UCDs. 
The observations are complemented by our new X-ray measurement for \denis. 
For the sake of enlarging the database of UCD activity, 
we also add unpublished X-ray observations of two UCDs that have in the past 
been observed in the radio band, 
\pcstar and \denisstar, both plotted in green in Fig.~\ref{fig:lr_lx}. 
We describe these targets, the X-ray observations and their analysis in the Appendix. 
Both objects are undetected in X-rays according to our analysis and in 
radio \citep[see][]{Berger02.1}.
As seen in Fig.~\ref{fig:lr_lx}, their upper limits do not place strong constraints. 

The most evident feature in Fig.~\ref{fig:lr_lx} is 
the violation of the Benz-G\"udel relation, first pointed out by \cite{Berger05.1}.  
Secondly, we notice a dichotomy for UCDs in terms of their radio and X-ray emission:
(A) Red in Fig.~\ref{fig:lr_lx} 
is a group of objects with X-ray flares and persistent X-ray emission but no radio emission
comprising LP\,412-31, LHS\,2065, vB\,10, Gl\,569\,B, and vB\,8. 
(B) Orange in Fig.~\ref{fig:lr_lx} 
is a group of objects with strong radio flares, mostly also with detected quiescent radio emission
but no or very weak X-rays (\tvlm, \tm, \lsr and possibly \bri). 
\denis (shown in green) also belongs to this latter group. 
LP944-20 is the only UCD for which flaring was observed both in the radio and in the X-ray bands.
For all other objects no flares have been recorded so far in X-rays or radio, most of them have not
been detected at either wavelength, and consequently they cannot 
(yet) be assigned to any of the two groups. 

The characteristics of the radio emission of group\,B has been shown to be very different from
that of G, K, and early-M dwarfs. 
While their quiescent emission may be ascribed to gyrosynchrotron radiation
as in GKM stars, a further, pulsed emission component which is 
highly circularly polarized has been found -- 
\tvlm \citep{Hallinan06.1}, \tm \citep{Berger05.1}, \lsr \citep{Hallinan08.1},  
and \denis \citep{Burgasser05.1} -- 
overlaid on their quiescent, non-variable emission. 
The periodicities of the spikes are 
consistent with the rotation rate of the objects implying an origin in a beaming mechanism. 
This radio component has been shown to be consistent with 
ECM instability. High inclinations were derived for
\tvlm, \tm and \lsr, leading \cite{Hallinan08.1} to ascribe the radio detection
of the ECM emission in these UCDs to a geometric selection effect. If this
scenario is true, high inclination is expected also for \denis, an hypothesis to be 
examined by means of photometric monitoring search for its rotation period.  

Group\,A could in principle be composed of UCDs in which the viewing conditions do not allow
us to observe the ECM pulses. However, the quiescent X-ray luminosities of its members are higher
than those for Group\,B objects. It is also striking that no coherent radio emitter is a strong and 
flaring X-ray source. 
Conversely, no flaring X-ray emitter is a (strong) radio source.  
(The exception is LP\,944-20 for which the ECM mechanism has not been 
explicitly confirmed.) This leads us to search for a possible origin of the apparent 
differences. 

\begin{figure}[t]
\begin{center}
\includegraphics[width=9cm]{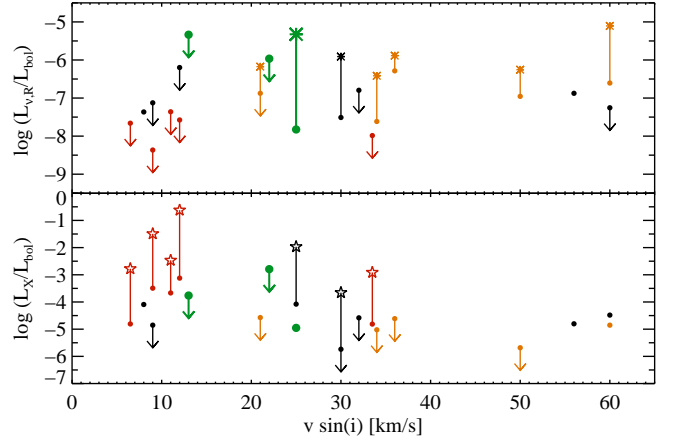}
\caption{Fractional radio and X-ray luminosity of UCDs versus rotational velocity; same plotting symbols as in Fig.~\ref{fig:lr_lx}.}  
\label{fig:vsini_LrandLx}
\end{center}
\end{figure}

All group\,A objects except Gl\,569\,B are in the sample for which \cite{Reiners07.2} have searched for magnetic fields, and strong fields between $1.3...3.9$\,kG were detected in all of them. 
No direct magnetic field measurements are available yet for the group\,B objects
with the exception of \denis \citep[$2.3 \pm 0.4$\,kG; ][]{Reiners10.0}. 
This observational bias is related to the fact that the ability to measure fields depends on the
rotation speed. Fast rotation makes it more difficult to detect magnetic line broadening. 
Fig.~\ref{fig:vsini_LrandLx} shows that stars in Group\,A are 
slower rotators than those in Group\,B.
The only fast rotator in Group\,A, Gl\,569\,B, has a large uncertainty in its $v \sin{i}$ value.
First, \cite{ZapateroOsorio04.1} give an uncertainty of $\sim 50$\,\% on their measurement of $v \sin{i}$. 
Secondly, Gl\,569\,B is a spectroscopic binary and we have plotted the mean of the values measured for the
two BDs ($30 \pm 15$ and $37 \pm 15$\,km/s for Gl\,569Ba and Gl\,569\,Bb, respectively). 
Two further UCDs have exhibited X-ray flares, 
SCR\,1845-6357 \citep{Robrade10.1} and RXS J115928.5-524717 \citep{Robrade09.2}; 
black with asterisks for flares in Fig.~\ref{fig:vsini_LrandLx}. While the former
one has a very low limit for its quiescent X-ray emission, the latter one 
has a quiescent X-ray detection, i.e. it shows similar X-ray behavior to Group\,A. 
Both objects have not yet been observed in the radio band. 

It was recognized already by \cite{Berger08.1} that the radio and the X-ray emission of UCDs 
show opposite trends with high rotational velocity and that fast rotators 
violate the Benz-G\"udel relation more severely than slow rotators. Their conclusion was that 
magnetic activity (as measured in the radio band) is regulated by rotation, 
and that the decline of X-ray emission for fast rotators is ruled by secondary effects like inefficient
plasma heating or coronal stripping. However, the X-ray detection of the fastest rotators, 
Kelu\,1 and \tvlm, both with $v \sin{i} = 60$\,km/s according to 
\cite{Mohanty03.1},
argues against a suppression of X-ray emission as a result of rotation. 
Kelu\,1 (spectral type L) is also one of the coolest objects in the sample. 
A possible explanation for the (apparent) absence of radio pulses in Group\,A, besides the inclination
effect put forth by \cite{Hallinan08.1}, is another selection effect related to their longer
rotation periods. For $v \sin{i} \sim 10$\,km/s and a `canonical' radius of $0.1\,R_\odot$ the
maximum period is $\sim 0.5$\,d, much longer than the typical duration of the radio observations
(typically $1-3$\,h). In most of the radio detected UCDs for which specific 
radio monitoring programs were undertaken a pulsed component was seen. The exception is
vB\,10 which did not show radio bursts in $10$\,h of continuous monitoring with the VLA
\citep{Berger08.1}.
However, with its low rotation of $6.5$\,km/s, the rotation period may be as high as
$0.8$\,d, almost twice as long as the radio observation.

\section{Summary and Conclusions}\label{sect:summary}

We have presented the first X-ray detection of the M9 dwarf \denis 
($\log{L_{\rm x}}\,{\rm [erg/s]} = 25.1$) 
and a broad-band spectrum from the UV to the NIR from which we redetermine
its stellar parameters 
(spectral type = M9, $T_{\rm eff} = 2450 \pm 50$\,K, $\log{g} = 4.3 \pm 0.3$) and 
kinematics ($RV = -11.4 \pm 2.0$\,km/s; 
$UVW$ space motions characteristic of the young disk).

The flux-calibrated broad-band spectrum together with the X-ray data 
allow us to extend flux-flux relations between chromospheric and coronal emissions 
into the regime of UCDs. 
We use the dwarfs of early- to mid-M type from \cite{MartinezArnaiz11.1} 
as comparison sample
for which we construct flux-flux relations between \ion{Ca}{ii}\,H+K,  Balmer
lines and soft X-rays. The correlations differ slightly from those 
determined by MA11 in an analogous study that included stars in a broader
range of spectral types (F-M). 
We compiled data from the literature for other UCDs. 
Given the large range over which the relations must be extrapolated from the
early-M dwarfs to the UCDs including \denis, the agreement of these objects with the
flux-flux relations is reasonably good. 
Discrepancies regard for \denis an excess of H$\alpha$ and X-ray
emission with respect to \ion{Ca}{ii}. Calcium measurements for other UCDs
are required to verify if this is a typical characteristic of UCDs.  

We have examined the Balmer decrements for \denis up to H8. 
A trend of increasing Balmer decrement towards later spectral type within
the M class, identified in a SDSS sample by BWHC07,  
is confirmed by our study in two ways: 
(1) We present the first measurement for M9 dwarfs (\denis and LHS\,2065).
Both objects show the highest H$\alpha$/H$\beta$ decrements measured so far 
within the M class. (2) The decrements $f_{\rm n}/f_{\beta}$ with $n=3...6$ 
are lower for early-M dwarfs (MA11 sample) than for late-M dwarfs 
(BWHC07 sample). 
However, contradicting this trend, 
the higher-n Balmer decrements of \denis are compatible with those
of the early M dwarfs from MA11 rather than with those of the (on average) later M dwarfs
from BWHC07. 

The observed Balmer decrements of \denis can be reproduced by an optically thick
low-temperature ($2500$\,K) plasma in LTE conditions or by an optically thin 
high-temperature ($20000$\,K) plasma in LTE conditions. 
The Case B recombination scenario is able to reproduce the
decrements Hn/H$\beta$ for $n=5$ and higher for a range electron densities between $10^{12}$ and 
$10^{14}\,{\rm cm^{-3}}$ and temperature $>10000$\,K  but these models can not fit the 
H$\alpha$/H$\beta$ ratio and the H$\alpha$/H$\beta$ ratio seems to be produced in a region of
lower density and temperature. 
For the densities and temperatures of the recombining plasma 
shown to describe the observed Balmer decrements from $n=5$ upwards 
one would also expect the detection of \ion{He}{I}\,D emission, contrary to our observation. 

The high H$\alpha$/H$\beta$ ratio of \denis is among the most curious findings of our study. 
It is this ratio that leaves doubts on the Case B interpretation, that leads to the 
low temperatures for the optically thick LTE scenario, and that is poorly
fitted by any optically thin LTE model.  
That such high H$\alpha$/H$\beta$ ratios may be realistic, and possibly even typical, 
for UCDs is supported by the results for vB\,10 and LHS\,2065 that we extracted from the literature. 
Nevertheless, we stress that the data for these two UCDs have very large or unknown
uncertainties and, therefore, the only firm measurements of the Balmer decrements in UCDs 
available so far are from our flux calibrated X-Shooter spectrum of \denis. 

Formally, the optically thick LTE emission provides the best match 
to the Balmer lines of \denis.
Comparing the measured line fluxes to the blackbody flux we estimate the line
emitting area to be between $2$\,\% (for H$\alpha$) and $100$\,\% (for H8) 
of the stellar surface, which can be considered reasonable values to the order of magnitude. 
We note that virtually all cases of Balmer line analyses treated in the literature refer to flares while 
for \denis we are dealing with non-flaring hydrogen emission. An exception where Balmer line fluxes are 
given for quiescent emission is the study of the M6 dwarf CN\,Leo by \cite{Fuhrmeister07.1}. 
For that case, the Balmer ratios H9/H$\beta$ to H15/H$\beta$ can not be described by optically thick
LTE plasma, and the lower Balmer lines are not available.
The low temperature of the optically thick LTE emission fitting the Balmer decrements of \denis 
suggests that the lines are emitted from just above the photosphere. 
This is at odds with the common notion that the hydrogen lines in active stars are produced in the
chromosphere at temperatures around ten-thousand Kelvin. 
The best-fitting optically thin LTE plasma provides such high temperatures. 
This model underpredicts the H$\alpha$ emission but, all in all, seems to be 
more plausible than the low-temperature optically thick case.

\denis clearly possesses a corona manifested by its X-ray detection, i.e. hot material
of MK temperatures is present. Further complicating the picture, the X-ray emitting plasma of \denis 
may not be associated with the radio
structures: \cite{Ravi11.0} have ascribed the gyrosynchrotron radio component to emission from 
electrons streaming out in open field lines that can not confine coronal plasma,
while the pulsed radio component represents ECM emission \citep{Burgasser05.1}, 
i.e. it has a physically different origin from the X-ray photons. 

Our X-ray detection of \denis also adds a new data point to the poorly sampled group of stars with
sensitive X-ray and radio observations. We tentatively identify two subgroups
of UCDs, X-ray flaring but radio faint objects with low rotation rate 
on the one hand and radio bursting but X-ray faint fast rotators 
on the other hand. \denis belongs to the latter class. At present it
remains unclear if this distinction corresponds to two physical populations
or if it is produced by observational biases.

\appendix

\section{New X-ray observations of UCDs}

\subsection{\pcstar}

\pcstar was identified by \cite{Schneider91.1} as an M star with outstanding
H$\alpha$ emission. On the basis of the lithium test \cite{Martin99.3} concluded
that this M9.5 object is a BD with age $< 1$\,Gyr. Variable veiling was ascribed to
activity, given the absence of NIR excess that would indicate an accretion disk. 
What makes \pcstar particular among the UCDs is that its 
very high levels of H$\alpha$ activity are persistent without evidence for flaring. 

\pcstar was observed with {\em Chandra} ACIS-S on 7\,Dec\,2001 
for $66$\,ksec (Obs-ID\,2579).
We analysed the data with standard CIAO\footnote{CIAO is the {\em Chandra} Interactive Analysis of Observations package provided by the {\em Chandra} X-ray Center (http://cxc.harvard.edu/ciao/index.html).} tools.
A description of the data reduction steps and calculation of upper limits 
can be found e.g. in \cite{Stelzer10.0}. 

No X-ray source is associated with \pcstar. We derive a $95$\,\% confidence level for 
the count rate upper limit of $9.6 \cdot 10^{-5}$\,cts/s in the $0.3-8$\,keV band. 
Assuming a one-temperature thermal spectrum of $\log{T}\,{\rm [K]} = 6.85$, 
PIMMS yields the count-to-flux conversion factor, 
that, when combined with the upper limit count rate, gives the flux limit. 
The distance of \pcstar has been reported to be $72$\,pc \citep{Dahn02.1}, 
and we obtain $\log{L_{\rm x}}\,{\rm [erg/s]} < 26.2$, 
one order of magnitude below the previous upper limit from {\em ROSAT}
\citep{Neuhaeuser99.1}. 
With a bolometric luminosity of $\log{(L_{\rm bol}/L_\odot)} = -3.62$ \citep{Dahn02.1}, 
the fractional X-ray luminosity is $\log{(L_{\rm x}/L_{\rm bol})} < -3.8$.

\subsection{\denisstar}

\denisstar stood out among the first stars identified by DENIS as being relatively
young. An upper limit to the age of $1$\,Gyr and a maximum mass of $55\,M_{\rm jup}$ 
was inferred from its lithium absorption \citep{Martin97.2}. \denisstar is 
classified as L5 spectral type \citep{Koerner99.1}. It was resolved into a 
nearly equal-mass binary \citep{Martin99.5, Brandner04.1} with separation of
$\sim 0.28^{\prime\prime}$, i.e. unresolvable by any present-day X-ray instrument. 

An archived {\em XMM-Newton} observation (ObsID 0002740601 and ObsID 0002740301) 
comprises $5.7$\,ksec of EPIC/pn exposure and $11.3$\,ksec of EPIC/MOS exposure. 
We analyzed the EPIC/pn 
data with standard {\em XMM-Newton} SAS\footnote{Science Analysis System for {\rm XMM-Newton} data reduction} tools. 
The highest sensitivity could be
obtained by merging the data from the three instruments. However, the unknown
relative sensitivity introduces a significant uncertainty. 
EPIC/MOS is typically three times less sensitive than EPIC/pn, but depending on 
the source spectrum. 
Therefore, despite the shorter exposure time, we concentrate on EPIC/pn data only. 

\denisstar is not detected in this X-ray observation. Analogous to the case of
\pcstar we derive a $95$\,\% confidence upper limit for the count rate ($0.007$\,cts/s)
and using the same assumptions for the spectral shape we obtain the flux limit.
The trigonometric parallax of \denisstar gives a distance of  
$20.2$\,pc \citep{Dahn02.1}, 
yielding an X-ray luminosity of $\log{L_{\rm x}}\,{\rm [erg/s]} < 26.6$ 
and with $\log{(L_{\rm bol}/L_\odot)} = -4.19$ \citep{Dahn02.1} the
fractional X-ray luminosity is $\log{(L_{\rm x}/L_{\rm bol})} < -2.8$.

\begin{acknowledgements}
We would like to recognize valuable input from an anonymous referee.
BS wishes to thank Fabrizio Bocchino for hot discussions on cool stars. 
BS acknowledges financial contribution from the agreement ASI-INAF I/009/10/0. 
KB acknowledges financial support from the INAF Postdoctoral fellowship.
This work is based on observations obtained with XMM-Newton, an ESA science mission with instruments and 
contributions directly funded by ESA Member States and NASA, 
and on observations made with ESO Telescopes at the VLT on the Paranal Observatory 
under programme ID $<$085.C-0238$>$. 
\end{acknowledgements}

\bibliographystyle{aa} 
\bibliography{denis1048}

\end{document}